\documentclass[fleqn,usenatbib]{mnras}
\usepackage{lmodern}

\usepackage{anyfontsize}
\usepackage{newtxtext,newtxmath}
\usepackage{graphicx}
\usepackage{subcaption} 
\allowdisplaybreaks
\usepackage[T1]{fontenc}
\usepackage[normalem]{ulem}

\DeclareRobustCommand{\VAN}[3]{#2}
\let\VANthebibliography\thebibliography
\def\thebibliography{\DeclareRobustCommand{\VAN}[3]{##3}\VANthebibliography}

\usepackage{graphicx}	
\usepackage{amsmath}	
\usepackage[dvipsnames]{xcolor}
\usepackage{multirow}

\newcommand{\vell}{{\vec{\ell}}}

\newcommand{\nv}{\hat{\bf n}}

\title[Constraints from Marked Power Spectra with HSC-Y1]{First Constraints from Marked Angular Power Spectra with Subaru Hyper
Suprime-Cam Survey First-Year Data}

\author[J. A. Cowell et al.]{Jessica A Cowell $^{1,2,3}$\thanks{E-mail: jessica.cowell@physics.ox.ac.uk}, {Joaquin Armijo$^{2,3}$}, {Leander Thiele$^{2,3}$}, {Gabriela A. Marques$^{4,5,6}$}, Camila P. Novaes$^{7}$,  \newauthor Daniela Grandón$^{8,9}$,  
{Sihao Cheng$^{10,11}$}, {Masato Shirasaki$^{12,13}$},  David Alonso$^{1}$, {Jia Liu$^{2,3}$} \\
$^{1}$Department of Physics, University of Oxford, Denys Wilkinson Building, Keble Road, Oxford OX1 3RH, United Kingdom\\
 $^{2}$Kavli IPMU (WPI), UTIAS, The University of Tokyo, 5-1-5 Kashiwanoha, Kashiwa, Chiba 277-8583, Japan\\
 $^{3}$Center for Data-Driven Discovery, Kavli IPMU (WPI), UTIAS, The University of Tokyo, Kashiwa, Chiba 277-8583, Japan\\
  $^{4}$ Centro Brasileiro de Pesquisas Físicas, R. Dr. Xavier Sigaud, 150 - Botafogo, Rio de Janeiro - RJ, 22290-180, Brazil\\
 $^{5}$Fermi National Accelerator Laboratory, P. O. Box 500, Batavia, IL 60510, USA\\
 $^{6}$Kavli Institute for Cosmological Physics, University of Chicago, Chicago, IL 60637, USA\\
 $^{7}$Instituto Nacional de Pesquisas Espaciais, Av. dos Astronautas 1758, Jardim da Granja, S\~ao Jos\'e dos Campos, SP, Brazil\\
 $^{8}$Instituto de Física y Astronomía, Facultad de Ciencias, Universidad de Valparaíso, Avenida Gran Bretaña 1111, Valparaíso, Chile\\
$^{9}$Mathematical Institute, Leiden University, Gorleaus Building,
Einsteinweg 55, NL-2333 CA Leiden\\
 $^{10}$Institute for Advanced Study, 1 Einstein Dr., Princeton, NJ 08540, USA\\
 $^{11}$Perimeter Institute for Theoretical Physics, 31 Caroline St N, Waterloo, ON N2L 2Y5, Canada\\
 $^{12}$National Astronomical Observatory of Japan (NAOJ), National Institutes of Natural Sciences, Osawa, Mitaka, Tokyo 181-8588, Japan\\
 $^{13}$The Institute of Statistical Mathematics, Tachikawa, Tokyo 190-8562, Japan
 \\
 }

\date{Accepted XXX. Received YYY; in original form ZZZ}

\pubyear{\the\year{}}

\begin{document}
\label{firstpage}
\pagerange{\pageref{firstpage}--\pageref{lastpage}}
\maketitle
\begin{abstract}
{We present the first application of marked angular power spectra to weak lensing data, using maps from the Subaru Hyper Suprime-Cam Year 1 (HSC-Y1) survey. Marked convergence fields, constructed by weighting the convergence field with non-linear functions of its smoothed version, are designed to encode higher-order information while remaining computationally tractable. Using simulations tailored to the HSC-Y1 data, we test three mark functions that up- or down-weight different density environments. Our results show that combining multiple types of marked auto- and cross-spectra improves constraints on the clustering amplitude parameter $S_8\equiv\sigma_8\sqrt{\Omega_{\rm m}/0.3}$ by $\approx$43\% compared to standard two-point power spectra. When applied to the HSC-Y1 data, this translates into a constraint on $S_8 =  0.807\pm 0.024$. We assess the sensitivity of the marked power spectra to systematics, including baryonic effects, intrinsic alignment, photometric redshifts, and multiplicative shear bias. These results demonstrate the promise of marked statistics as a practical and powerful tool for extracting non-Gaussian information from weak lensing surveys.}
\end{abstract}




\section{Introduction}
  A key challenge faced by existing and upcoming surveys is to extract cosmological information from the complex, non-Gaussian matter field formed through non-linear gravitational evolution. One of the most promising means to study the matter inhomogeneities is weak gravitational lensing, the subtle distortion of light paths caused by the gravitational tidal field of the large-scale structure~\citep[see recent reviews by e.g.][]{2015Kilbinger, Mandelbaum2018}. The lensing convergence field traces the integrated mass density along the line of sight and is therefore sensitive to the fundamental cosmological parameters encoded within it.

  Most lensing analyses, \citep[for example]{DESY3_2022, Dalal_2023, 2025kidslegacy}, rely on two-point statistics, such as the two-point correlation function or the power spectrum,  which measure matter clustering as a function of scale. They are optimal for Gaussian fields, whose statistical information is fully captured in the covariance matrix. However, the late-time matter density field is strongly non-Gaussian, motivating the development of new statistics that can recover the non-Gaussian (NG) information missed by two-point statistics, especially for use in future surveys with more precision~\citep[e.g.][]{EuclidHOWLS2023}. Examples of NG statistics relevant for weak lensing include but are not limited to:  higher order moments \citep{Secco_2022,gatti2024darkenergysurveyyear}, peaks and minima counts \citep{liu2015,Martinet_2018,Zürcher_2022, marques2023cosmologyweaklensingpeaks, 2025MNRAS.536.2064G}, Minkowski functionals \citep{2015PhRvD..91j3511P}, scattering transforms \citep{ gatti2024darkenergysurveyyear}, persistent homology \citep{Calles_2024}, probability distribution functions \citep{liu2019,Boyle2021,thiele2023cosmologicalconstraintshscy1}, machine learning \citep{2022Fluri}, field level inference \citep{Zhou_Li_Dodelson_Mandelbaum_2024} among others.

  Alternatively, instead of creating new statistics, one could perform a non-linear transformation on the fields before applying the usual two-point statistical machinery. By doing so, higher-order information is brought into the statistic.  Our work adopts one such transformation, known as \textit{marking}--also referred to as marked statistics, the marked power spectra, or the marked correlation function. This method was first developed for the galaxy overdensity field, where galaxies were weighted by their intrinsic properties before their spatial correlations were computed~\citep{1984Stoyan, Sheth2005Conolly,  Sheth_2005, Skibba_2006,2021A&A...653A..35S}. More recently, density-marked statistics, where the field is weighted by a non-linear function of the smoothed density field, have been introduced to enhance constraints on modified gravity \citep{White_2016,Valogiannis_2018,Armijo_2018,Hernández-Aguayo_Baugh_Li_2018,  kärcher2024optimal,  Armijo_2024a, Armijo_2024b}, neutrino masses \citep{Massara_2021}, and other cosmological parameters \citep{Satpathy_2019,  Philcox_2020_pert,Aviles_2020,  2021Philcox,  Massara_2023, Cowell2024, Marinucci_2024, Ebina}. The marked power spectrum was recently applied to galaxy clustering data and achieved $1.2\times$ tighter constraints on $S_8\equiv\sigma_8\sqrt{\Omega_{\rm m}/0.3}$~\citep{2024SimbigMarks}. Although analytical approaches to the marked power spectrum have been developed using perturbation theory \citep{ Philcox_2020,kärcher2024optimal, Ebina}, they break down at small scales. Simulations offer a natural alternative, enabling accurate modelling of non-linear regimes beyond the reach of perturbative methods. For example, \cite{Cowell2024} employed Fisher analysis to optimise information extraction using $N$-body simulations with and a Gaussian process–based mark function. Simulations also allow us to incorporate small-scale observational effects and survey geometry, which are difficult to model analytically.

  In this work, we apply the marked power spectrum for the first time to weak lensing data, using convergence maps derived from the HSC-Y1 survey. This also represents the first application of marked statistics to projected fields. This paper is complemented by a series of papers applying non-Gaussian statistics to HSC-Y1 using the same observational data and simulations that investigate counts of peaks and minima counts \citep{marques2023cosmologyweaklensingpeaks}, probability distribution functions (PDFs),\citep{thiele2023cosmologicalconstraintshscy1}, scattering transform coefficients  \citep{Cheng2025}, Minkowski functionals \citep{Armijo:2024ujo}, impact of baryonic feedback on NG statistics \citep{grandon2024impactbaryonicfeedbackhsc}, and a combination of NG statistics with a likelihood-free inference approach~\citep{novaes2024cosmologyhscy1weak}. 

  The structure of this paper is as follows: in Section \ref{sec:theory} we describe the theory of marked fields, and our choices of mark functions. In Section \ref{sec:methods}, we discuss the data, simulations, emulation methods, and systematic tests. We present our results in Section \ref{sec:results} and the main conclusions of this work in Section \ref{sec:conc}.

\section{Theory of Marked Fields}\label{sec:theory}
  \subsection{Mark Functions}
    Density-weighted marks have traditionally been used for three-dimensional~(3D) density fields, such as the matter or galaxy fields. In this paper, we follow the same spirit but apply marking to the 2D lensing convergence field $\kappa$, which is proportional to the matter overdensity integrated along the line of sight. We define the convergence-weighted marked field $\Delta(\kappa)$ as,
    \begin{equation}
      \Delta(\kappa) = m(\kappa_\theta) \kappa,
      \label{eq: kappa_mark_eqn}
    \end{equation}
    where $m(\kappa_\theta)$ is the mark function and $\kappa_\theta$ is the convergence field smoothed with a Gaussian filter of width $\theta$.  In this work, we broadly refer to the angular power spectrum of the marked convergence field $C_\ell^{\Delta\Delta}$, the cross-spectrum with the original $\kappa$ field $C_\ell^{\Delta\kappa}$, as well as their combination as the marked power spectrum, or spectra in the case of multiple mark functions.

  \subsection{Choice of Mark Function}\label{ssec:choice of mark}
    The choice of the mark function remains an open question. While \cite{Cowell2024} looked to find the optimal mark for cosmological parameters $\Omega_{\rm m}$ and $S_8$ using $N$-body simulations at a set redshift, the choice of the function maximising the information gain may be vastly different depending on the targeted parameter, or even properties of the field itself, such as redshift. For example, applications to studies of modified gravity \citep{kärcher2024optimal}, neutrino mass \citep{Massara_2021}, and primordial non-Gaussianity \citep{Marinucci_2024}, would likely need to probe different volumes and scales, and therefore up-weight different regions. Moreover, the choice can be different if one wants to limit oneself to an analytically tractable mark \citep{Ebina, Marinucci_2024}, or explore options computationally. Moreover, as shown in \citet{Cowell2024}, if all auto- and cross-spectra are used, the mark function has an affine symmetry, which allows one to transform two seemingly different marks into one another and still get equivalent constraints.

\begin{figure*}
      \includegraphics[width=\linewidth]{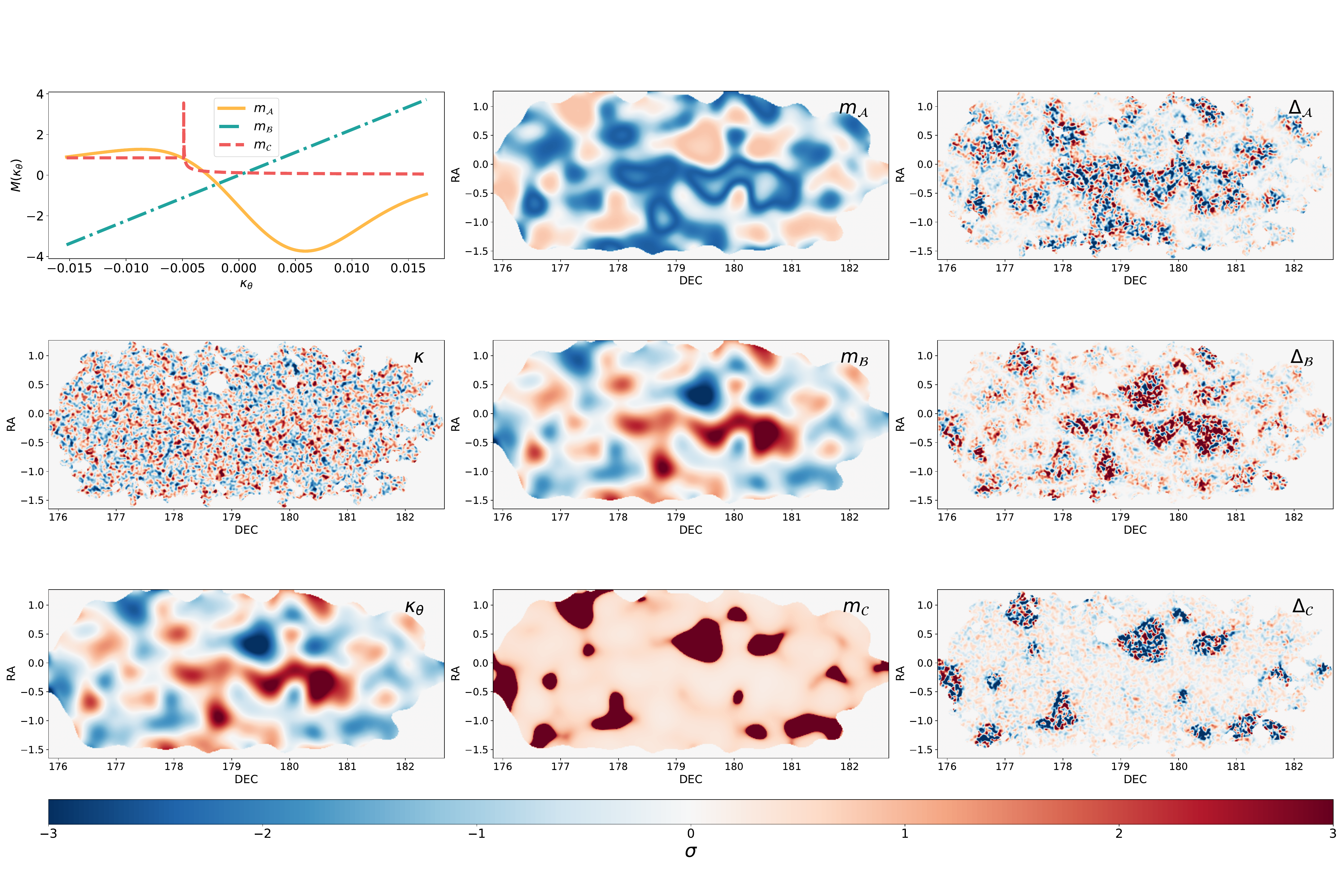}
      \caption{We illustrate the process of creating marked fields $\Delta_i$, using the example of the convergence field from the HSC-Y1 ``wide12h'' field with a $10'$ smoothing scale in the third redshift bin ($0.9 < z < 1.2$).  Left column (top to bottom): example mark functions as a function of the smoothed field $m_{i}(\kappa_\theta)$; the original convergence field $\bf \kappa$, smoothed at $1'$; and the smoothed field $\kappa_\theta$ for $\theta = 10'$. Middle column: mark functions applied to $\kappa_\theta$. Right column: marked fields $\Delta_i = m_i(\kappa_\theta)\, \kappa$, used for calculating the marked power spectra. All mark functions and fields are renormalised by their standard deviations for visualisation. The colour thus represents fractions of the standard deviation for each field.}\label{fig: mark_examples}
    \end{figure*}

    As a first study on projected fields, we focus on the application, leaving the optimisation of the mark functions for future work. We adopt three distinct mark functions from the recent literature, each with a different weighting profile of the density field:
    \begin{enumerate}
      \item[($\mathcal{A}$)] The Gaussian process (GP) mark introduced in \citet{Cowell2024}, which we denote as $m_\mathcal{A}$. We use the `optimal' mark shape found for a 10 $h^{-1}$ Mpc smoothing scale, rescaling it, such that the 'nodes' are spaced equidistantly between the maximum and minimum values of the smoothed convergence field. This mark function is shown in yellow in the top left panel of Figure~\ref{fig: mark_examples}. This mark function is a smooth function between four predefined nodes and fitted by a GP -- a collection of multivariate normal random variables, fully defined by a mean function (which we set to zero) and a kernel function parametrising the covariance between any two points. For the kernel, we use the radial basis function, also known as the ``squared exponential'' kernel (SE), which is specified by only two parameters: an amplitude $\sigma^2$ and length scale $\ell$:
      \begin{equation}
        k_{\rm SE}(x,x')=\sigma^2\,\exp\left(-\frac{(x-x')^2}{2\ell^2}\right).
      \end{equation}
      Using this kernel\footnote{We set the amplitude to be 20 and the length scale to be 0.7. However, these choices are somewhat arbitrary -- so long as the length scale is larger than the distance between nodes, one can recover the shape of the mark function.}, the mark function $m(\kappa_\theta)$ is defined in terms of its values at a set of four nodes at fixed $\kappa_\theta$ values. We label the node positions $\boldsymbol{\kappa}_\theta^*\equiv(\kappa^*_{\theta,1},\cdots,\kappa^*_{\theta,4})$ with corresponding values of mark function ${\boldsymbol m}_*\equiv(m(\kappa_{\theta,1}^*),\cdots,m(\kappa_{\theta,4}^*))$. The value of the function at any other $\kappa_\theta$ is given by the most likely GP conditioned on the known node values:
      \begin{equation}
        m_\mathcal{A}(\kappa_\theta)={\bf k}^T_*(\kappa_\theta)\,{\bf K}^{-1}_*\,{\bf m}_*,
      \end{equation}
      where the elements of the vector ${\bf k}_*(\kappa_\theta)$ and matrix ${\bf K}_*$ are
      \begin{equation}
        {\mathrm k}_{*,i}(\kappa_\theta)\equiv k_{\rm SE}(\kappa_\theta,\kappa^*_{\theta,i}),\hspace{12pt}
        {\mathrm K}_{*,ij}\equiv k_{\rm SE}(\kappa^*_{\theta,i},\kappa^*_{\theta,j}).
      \end{equation}
      The $\kappa$ position of the 4 nodes is recalibrated to be equally spaced between the maximum and minimum values of each field, incorporating the different ranges of the $\kappa$ fields, capturing the dependence on smoothing scales, galaxy shape noise, and redshifts. By doing this, we are effectively introducing slightly different mark functions while preserving the general shape of the function.\\
 
      \item[($\mathcal{B}$)] The mark given by the smoothed field itself re-weighted by its standard deviation for stability, which we denote as $m_\mathcal{B}$ and show in green in the top left panel of Figure~\ref{fig: mark_examples}. i.e. 
      \begin{equation}
        m_{\mathcal{B}} = \frac{\kappa_\theta}{\sigma(\kappa_\theta)}.
      \end{equation}
      This choice is inspired by the recent work by \cite{Ebina}, exploiting the fact that this lower-order mark is analytically tractable. It is also intuitive to see the contributions from higher-order terms arising in this mark, for example, the correlation function of the original field and marked field, i.e. $\langle \kappa(\nv)\,\Delta(\nv')\rangle=\langle\kappa(\nv)\, \kappa(\nv')\,\kappa_\theta(\nv') \rangle /\sigma(\kappa_\theta)$. We expect this to be similar to skew- or kurt-spectra \citep{Munshi_2020, Munshi_Lee_Dvorkin_McEwen_2022}, with a difference introduced by the smoothing kernel, which we discuss later. \\

      \item[($\mathcal{C}$)] A modified version of the mark function originally introduced in \citet{White_2016} in the context of modified gravity, which we denote as $m_\mathcal{C}$ and show in red in the top left panel of Figure~\ref{fig: mark_examples},
      \begin{equation}   
        m_{\mathcal{C}}(\kappa_\theta)= (1 + f(x))^{-p},
        \label{eq:white}
      \end{equation}
      where
      \begin{equation}
        x = \frac{\kappa_\theta}{\sigma(\kappa_\theta)(1 + b )},
      \end{equation}
      with the safety function $f$ defined as
      \begin{equation}    
        f(x) =
        \begin{cases}
          x, & \text{if } x \geq -0.999 \\ 
          (1 - \delta) \tanh(10x), & \text{otherwise} 
        \end{cases}
        \label{eq:safety func}
      \end{equation}
      to avoid singularities. In this analysis, we choose the values $p=0.5$, $\delta=0.02$, and $b=0.1$ for the free parameters of this mark function, closely following the parameters used by \citet{Massara_2023, Armijo_2024b}. 
      This mark function up-weights under-dense regions of the field. Compared to the negative power-law up-weighting adopted by~\cite{White_2016}, our function up-weights underdense regions with a constant factor, which is more suitable for 2D projected fields. 

    \end{enumerate}
 
     We show an example of the mark functions and their fields for an example HSC-Y1 sky patch in Figure \ref{fig: mark_examples}. In the left column, we show the three marks as a function of the convergence value in the upper panel, the original convergence field $\kappa$ in the middle panel, and its smoothed version with a smoothing scale $\theta=10$ arcmin in the lower panel. In the middle column, we show the marks $m_i({\bf\kappa}_\theta)$ for the three mark functions. In the right column, we show the corresponding marked fields $\Delta_{i}$ (Eq.~\ref{eq: kappa_mark_eqn}), from which we compute the marked power spectrum. 

    We can see the intuitive motivation of each mark more clearly in these plots, by looking at the bright regions in the middle and right columns, and where they correspond to in the original $\kappa$ and $\kappa_\theta$ maps in the left column. The GP mark, $m_\mathcal{A}$, anti-correlates low-medium overdensities and underdensities, such that the marked field appears inverted when compared with $\kappa_\theta$. Meanwhile, $m_\mathcal{B}$ is proportional to the smoothed field, enhancing the extreme overdensities and underdensities.  $m_\mathcal{C}$ up-weights under-dense regions, but is equal to a constant for positive $\kappa$, such that we expect the information content from these regions of this marked field to have no additional information compared to the standard power spectra. Note that this example is for 10 arcmin smoothing, but we also include down to 2 arcmin smoothing in the baseline analysis, where the small-scale structure and scale of the bright regions are more comparable. 

\section{Methods}\label{sec:methods}
  \subsection{HSC-Y1 Data}
    We use lensing convergence maps estimated from the HSC-Y1 galaxy shapes catalogue \citep{Mandelbaum_2017}. The data consists of 6 fields, spanning 136.9 deg$^2$, and {source} redshifts $ 0.3 < z < 1.5$, calculated using the MLZ code \citep{Tanaka_2017}. After masking, there is a total number density of 17 arcmin$^{-2}$ galaxies. Source galaxies are split into four tomographic redshift bins with edges $[0.3,\, 0.6,\,0.9,\,1.2,\,1.5]$. Convergence maps are constructed from the shear maps using the Kaiser-Squires (KS) inversion \citep{1993Kaiser}, which we repeat for smoothing scales of 1, 2, 4 and 10 arcmin. We also perform inpainting on the maps before performing the KS inversion to avoid E-B leakage due to the mask. Finally, we reapply the corresponding smoothed masks to the convergence fields.  In our analysis, we treat the 1 arcmin smoothed maps as the ``original'' $\kappa$ maps, and will denote them as $\kappa$ rather than $\kappa_{\theta=1}$.

  \subsection{Simulations}\label{ssec:sims}
    We use two sets of simulations, the ``covariance set'' and the ``cosmo-varied'' set. The covariance matrix is estimated from a set of 2268 pseudo-independent realisations generated from 108 full-sky $N$-body simulations. They are run with a fiducial cosmology taken from the best-fit result of the Wilkinson Microwave Anisotropy Probe nine-year data \citep{2013WMAP} ($\Omega_b =0.046, \Omega_{\rm m} = 0.279, \Omega_\Lambda=0.721, h = 0.7, \sigma_8 = 0.82,  n_s=0.97 $), generated from the simulations in \citep{Takahashi_2017}. To model the marked power spectra in lieu of an analytical prediction, we use a separate suite of 100 cosmology-varied simulations presented in \cite{Shirasaki_2021, marques2023cosmologyweaklensingpeaks}. These simulations vary $\Omega_{\rm m}$ and $\sigma_8$, with 50 quasi-independent realisations for each cosmological model.

    The produced mocks incorporate observational effects of the HSC-Y1 catalogue, such as multiplicative bias, inhomogeneity of source galaxies, survey geometry, variations in the lensing weight, redshift and galaxy distribution. Full details of the method for tailoring the simulations to the HSC-Y1 data as well as adding systematic effects to the simulations, can be found in \citet{marques2023cosmologyweaklensingpeaks,Shirasaki_Yoshida_2014, Shirasaki_Hamana_Takada_Takahashi_Miyatake_2019}. 

  \subsection{Power Spectra}
    The calculation of the marked or standard power spectra is identical, with the only difference being the field itself. We use the flat sky approximation to calculate the power spectra of the fields using the pseudo-$C_\ell$ approach \citep{Hivon_2002} as implemented in {\tt NaMaster}  \citep{Alonso_2019}, which corrects for the coupling of different Fourier modes due to the mask. We compute power spectra in 14 logarithmically spaced bins in the range $80 < \ell < 6500$, following \cite{Hikage_2019,marques2023cosmologyweaklensingpeaks}. We take the weighted average of the power spectra calculated for each of the 6 HSC-Y1 fields, where the weights are equal to the number of galaxies per field. 

  \subsection{Covariance}\label{sec:cov}
    \begin{figure*}
      \centering
      \includegraphics[width=\textwidth]{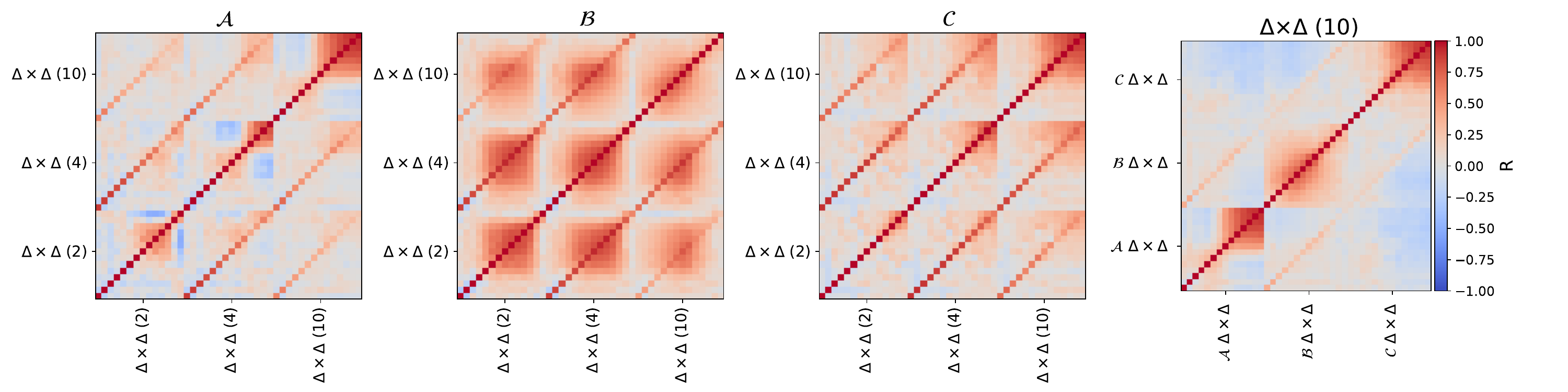}
      \caption{Example correlation matrices for sections of the full covariance matrix. In the first three columns, we show the blocks corresponding to the auto spectra ($C_\ell^{\Delta\Delta}$ of the three marks, $\mathcal{A,B,C}$, defined in Section \ref{ssec:choice of mark} for smoothing scales of $2'$, $4'$ and $10'$. On the far right, we show the correlation between these marks for the example of 10'. All blocks are before scale cuts. All are shown using the third redshift bin $(0.9 < z < 1.2)$ as an example. )}
      \label{fig:corr}
    \end{figure*}
    Our full data vector for the baseline analysis consists of the $C_\ell$ measurement of the original field $\kappa$ map smoothed to one arcminute, and the cross and auto spectra of nine marked fields, consisting of three different mark functions and three smoothing choices, $\theta=[2 ', 4 ', 10']$\footnote{Note this is the smoothing scale of the \textit{mark function}, such that a smoothing scale of $10'$ will still include information from the 1' field due to the definition of the mark function in Equation \ref{eq: kappa_mark_eqn}.} Our data vector will then be as follows, 
    \begin{equation}
      \mathbf{D} = \Big\{ C_\ell^{\kappa, \kappa}, \quad C_\ell^{\kappa, \Delta_{\mathcal{X},\theta}}, \quad C_\ell^{\Delta_{\mathcal{X}\theta}, \Delta_{\mathcal{X}\theta}} \Big\}
    \end{equation}
    where $\mathcal{X} \in \{\mathcal{A}, \mathcal{B}, \mathcal{C} \}$, $\theta \in \{2', 4', 10'\}$, and $\Delta_{\mathcal{X},\theta} \equiv m_\mathcal{X}(\kappa_\theta)\kappa$. Note that we do not include cross-correlations between separate marks, smoothing scales, or tomographic bins. We use three redshift bins, omitting the fourth bin due to potential issues with the photometric redshift calibration \citep{Marques_2024, Dalal_2023,Li_2023}, and 14 $\ell$ bins, leading to a data vector of length 1064 before scale cuts, or 380 after scale cuts (discussed in Section \ref{ssec:scale_cuts}).

    It is crucial to verify that the covariance matrix is accurately estimated, especially for such a large data vector. When inverting the covariance matrix, we apply the Anderson–Hartlap factor \citep{2007_Hartlap} to correct for the finite number of simulations used to calculate the covariance, such that the inverted covariance is $C^{-1}_{\rm corrected}= \alpha C^{-1}$, where $\alpha = \frac{N_s-N_d -2}{N_s -1} $, $N_d=380$ is the number of data-points, and $N_s=2268$ is the number of simulations.

    We examine the correlation matrix across marks and smoothing scale, and show an example subset for cross spectra in Figure~\ref{fig:corr}. The first three panels show correlations across smoothing scales for individual marks. The fourth panel shows correlations across the three mark functions. We only show data using the third redshift bin $(0.9 < z < 1.2)$ as an example. \( m_\mathcal{B} \) displays strong correlations across scales due to its construction (second panel). Negligible correlations between different marks are seen (fourth panel), demonstrating the relatively independent information they carry. This motivates our combined analysis of all three marks. 

  \subsection{Emulator}\label{ssec:emulator}
    We use a Gaussian process to emulate our summary statistics, i.e. the power spectrum and various marked power spectra. We use the GP Regression implemented in {\tt scikit-learn}\footnote{\url{https://scikit-learn.org/}} using a radial basis function kernel with a fixed length scale of 5. We found the emulator accuracy independent of the length scale. Each bandpower is emulated with its own GP Regression, such that points at different bandpowers are effectively treated as independent. The emulator is trained on 100 different cosmologies, using the average (marked) power spectrum over 50 realisations for each cosmology. We evaluate the emulator’s accuracy using cosmology-varied simulations in two stages, both employing the ``leave-one-out'' approach, where the emulator is trained on 99 out of 100 cosmological models, and then used to predict the power spectrum and corresponding marked power spectra for the excluded model. 

    First, we test the accuracy of the predicted power spectra and marked power spectra for the test model. For each cosmology, we calculate the difference between the predicted power spectra and the true power spectrum, 
    averaging over 50 realisations. For all scales, the emulator is accurate to within $\sim10\%$ standard deviation, where the standard deviation is evaluated across realisations. We also visually inspect these deviations to check that there is no trend with cosmology or scale.
    \begin{figure*}
      \centering
      \includegraphics[width=\linewidth]{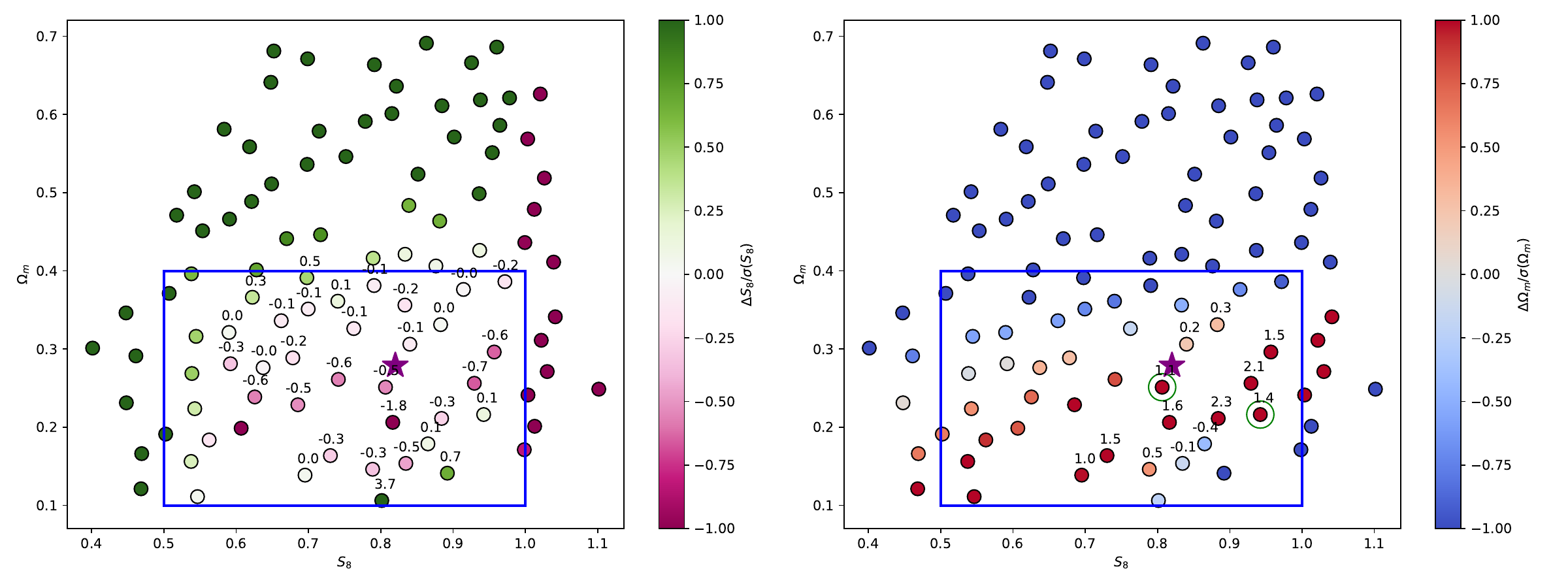}
      \caption{Accuracy validation using the leave-one-out method. For each of the 100 cosmologies (shown as points), we remove that simulation from the emulator and use it as an input data vector to evaluate the shift in the inferred values of $S_8$ (left) and $\Omega_{\rm m}$ (right) relative to the true values. Offsets are expressed in units of the posterior standard deviation. The analysis uses the baseline setup with all three mark functions and $\ell_{\rm max} = 1500$. The box indicates the prior range; numerical offsets are displayed only for posteriors that remain within this range. Contours shown in green correspond to cases with numerical biases exceeding $1\sigma$ but that do not appear concerning upon visual inspection.}
      \label{fig:post}
    \end{figure*}

    Next, we investigate the accuracy of the inferred cosmology when sampling with an emulated data vector. We show deviations in cosmological parameters $S_8$ and $\Omega_{\rm m}$ using the emulator (mean of the posterior), in comparison to the true value of the test cosmology, in Figure~\ref{fig:post}. Here we use the combination of all three marks. The deviations are quantified with respect to the standard deviation of the parameter posterior as calculated using the {\tt GetDist} package \citep{Lewis:2019}. The purple star corresponds to the cosmology of the covariance-training simulations, while the blue box indicates the prior regions used. The points outside the prior are still used to train the emulator. Interestingly, we identify one point located relatively close to the fiducial cosmology used for the covariance matrix estimation that shows a $2\sigma$ bias in $S_8$. Although the origin of this bias remains unclear, we verified that including or excluding this point from the emulator has no impact on our final constraint. Our analysis shows a large deviation on the $\Omega_{\rm m}$ value, which is largely due to the fact that $\Omega_{\rm m}$ is prior-dominated, giving an artificially smaller standard deviation. Therefore, in our final results, we focus on  $S_8$ instead of $\Omega_{\rm m}$. We refer the reader to Appendix \ref{sec:emulator_appendix} for more details.
  
  \subsection{Likelihood estimation}
    To reduce noise in the covariance matrix, we compress our data vector and covariance matrix using the {\tt moped} compression scheme \citep{Heavens_2000}. This reduces our data vector to only 2 elements, corresponding to the number of parameters to be inferred. We assume a Gaussian likelihood, 
    \begin{equation}
      \mathcal{L}({\bf{D}}|\theta_p) = -\frac{1}{2}\left[ {\bf D} -{\bf x}(\theta_p)\right]^T{\sf C}^{-1}[{\bf D -x}(\theta_p)] + {\rm const.},
    \end{equation}
    where {${\bf D}$ is the summary statistic of the data, ${\sf C}^{-1}$ is the inverse of the covariance matrix with the Anderson–Hartlap correction, and the theoretical prediction ${\bf x}(\theta_p)$} is produced by the emulator described in Section \ref{ssec:emulator}, as a function of the parameters $\theta_p = [\Omega_{\rm m}, S_8]$. We use the {\tt cobaya} MCMC sampler \citep{Torrado_2021, Lewis_2002} to sample the posterior, using 10 chains with uniform priors on $\Omega_{\rm m}$ of [0.1, 0.4] and $S_8$ of [0.5, 0.1].

  \subsection{Scale Cuts and Systematics}\label{ssec:scale_cuts}
    The marked power spectrum is constructed by weighting the local density field with a function of its smoothed counterpart in real space, which inherently mixes different scales in Fourier space. As a result, the scale cuts typically applied to the two-point correlation function may not be appropriate for marked power spectra. Similarly, various systematics may exhibit different scale dependencies compared to the standard power spectrum. We explore the full range of scales in the presence of systematics and provide a more detailed discussion in Appendix \ref{apec:systematics}. 

\begin{figure*}
      \centering
      \includegraphics[width=\linewidth]{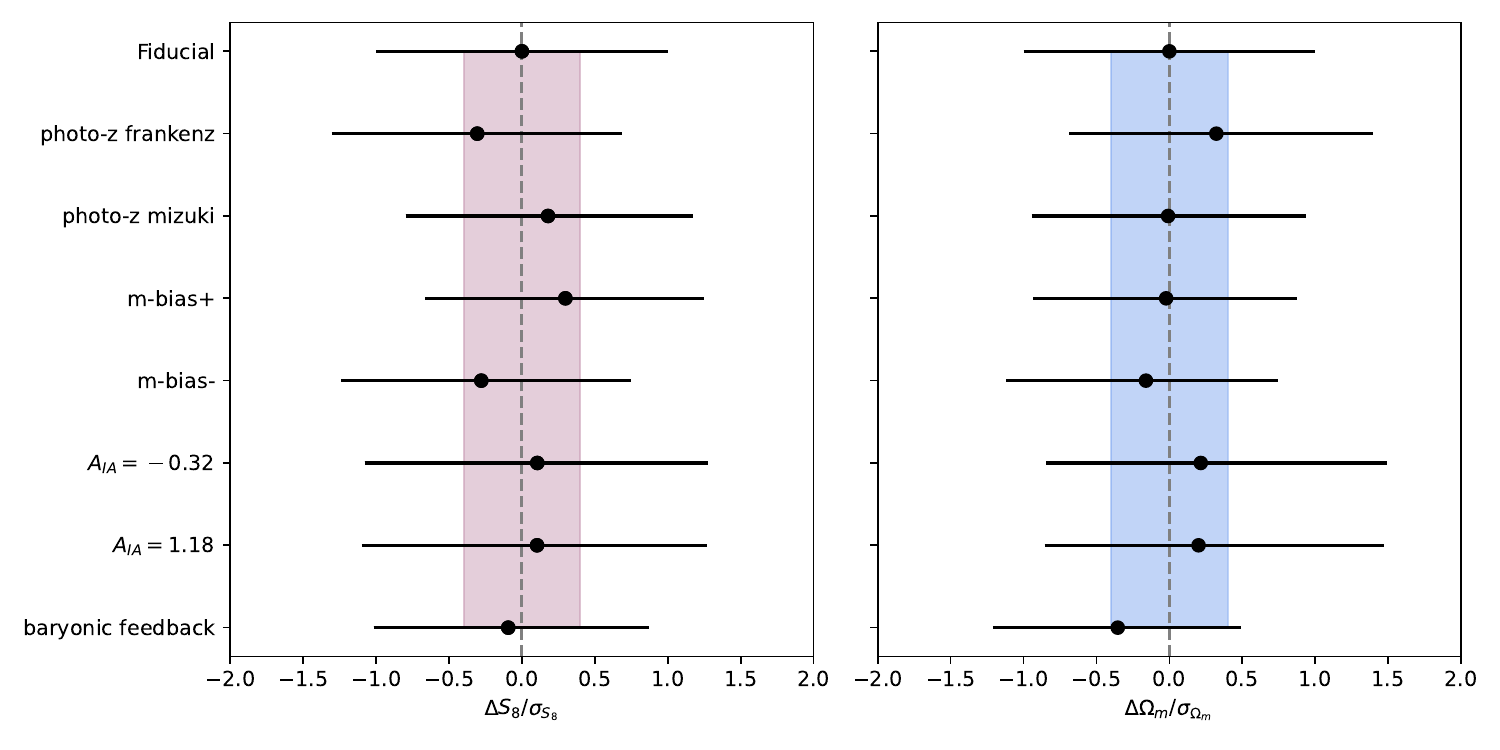}
      \caption{Analysis of the impact of systematic effects on $S_8$ (left) and $\Omega_{\rm m}$, (right). For each contaminated data vector we repeat the inference and recover the error bars shown. The coloured regions show $1/3\sigma$ confidence interval of the underlying cosmology with no systematics added. We check to see the systematics do not bias the inference outside of this range.}\label{fig:scale cuts}
    \end{figure*}

    For our baseline analysis, we include all three mark functions and three smoothing scales for the mark function, $\theta=[2', 4', 10']$. We select scale cuts such that the induced shift in $S_8$ due to systematics satisfies $\Delta S_8 \lessapprox 0.4\sigma(S_8)$ across all tested systematics. We find that adopting a maximum multipole of $\ell_{\text{max}} = 1500$ meets this criterion, and we present the corresponding inferred values from simulations with systematic contamination in Figure \ref{fig:scale cuts}. The top data point of the plot shows the ``Fiducial'' case, in which no systematics are present. The shaded bands represent the $\pm 0.4\sigma_{\rm Fid}$ thresholds for both $S_8$ and $\Omega_{\rm m}$. Below this, we show the shifts in inferred parameter values resulting from systematic-contaminated data vectors. We consider the analysis robust if these shifted values remain within the shaded bands. We adopt the same systematic tests employed in previous papers in this series, with a detailed description of the contamination procedures provided in \citet{marques2023cosmologyweaklensingpeaks}. For completeness, we also provide a summary of these procedures below.

  \subsubsection{Baryonic Effects}
    Baryonic effects are known to suppress the matter power spectrum on small scales. Since marked power spectra can transfer information between small and large scales, it is essential to test their sensitivity to such effects. To do this, we compute the marked power spectra using both hydrodynamic and dark matter-only simulations from the $\kappa$TNG dataset \citep{Osato_Liu_Haiman_2021}. We then contaminate the data vector, generated from the ``covariance training set'' simulations, using the ratio of the hydrodynamic to dark matter-only statistics, and examine the resulting parameter shifts. Overall, we find that baryonic effects introduce minimal biases in the $S_8$ inference, even when small scales are included, as further discussed in Appendix \ref{apec:systematics}. For our baseline scale cut choice, we see a shift of only $-0.09\sigma_{\rm Fid}(S_8)$. We find the bias on $\Omega_{\rm m}$ to be much higher, although still within the statistical uncertainties (approximately $-0.35 \sigma_{\rm Fid}(\Omega_{\rm m})$). A related study by \citet{grandon2024impactbaryonicfeedbackhsc} analysed the influence of baryonic feedback on peak counts, minimum counts, the probability distribution function, and the scattering transform in the HSC-Y1 dataset. They similarly reported negligible shifts in $S_8$ across all statistics. 
  
    \subsubsection{Intrinsic Alignment}
      Intrinsic alignments (IA) are a known source of systematic error in weak lensing measurements, arising when galaxy shapes are influenced by local interactions and become intrinsically correlated to the surrounding large-scale structure. Although many models exist to describe this effect, in this work we focus solely on the non-linear Tidal Alignment (NLA) model \citep{Bridle_2007}. In the NLA framework, the strength of intrinsic alignment is controlled by the coupling parameter $A_{IA}$, such that IA-matter and IA-IA power spectra are
      \begin{equation}\label{eq:IA}
        P_{Im}(k,z)=A_{IA}(z)\,P_{NL}(k,z)\hspace{12pt}
        P_{II}(k,z)=A_{IA}^2(z)\,P_{NL}(k,z),
      \end{equation}
      where $P_{\rm NL}(k,z)$ is the non-linear matter power spectrum.

      Following previous studies on the HSC-Y1 data, values were reported as $A_{IA} = 0.38 \pm 0.7$\citep{Hikage_2019}, and  $A_{IA} = 0.91^{+0.27}_{-0.32}$ \citep{Hamana__2020}. For this analysis, we choose to test two values of $A_{IA}=\{-0.32, 1.18 \}$, corresponding to the extremes of these cases within $1\sigma$. We find the intrinsic alignment to give biases of up to $0.12\sigma_{\rm Fid}(S_8)$.

    \subsubsection{Photometric redshifts}
      Photometric redshift (photo-z) estimation is an important challenge in cosmological weak lensing, and a potential source of the $S_8$ tension \citep{2025kidslegacy}. In HSC, the redshifts are determined using several independent codes, described in \cite{Tanaka_2017}. Our baseline analysis uses source redshifts ranging to $0.3 < z < 1.5$, and uses the best-fit photo-z determined by a machine-learning code based on self-organizing maps (MLZ). We test the effect of using two alternative redshift estimate techniques, using an extra 100 simulations generated with the Flexible Regression over Associated Neighbors with Kernel dEnsity estimatioN for Redshifts (FRANKEN-Z) and the Mizuki method \citep{Tanaka_2017} which utilised theoretical stellar population synthesis models, and using the power spectra from these simulations as input datavector. For the Mizuki method, we find a shift of $+0.2\sigma_{\rm Fid}(S_8)$ for the baseline analysis choices. However, we find the bias caused by redshift estimation to have an unpredictable behaviour with scale, as discussed in Appendix \ref{apec:systematics}. We found the FRANKEN-Z photo-z method to give the least predictable systematic shifts for the $S_8$ parameter out of all systematics. With more conservative scale cuts, $\ell<1000$, we find a shift of $-0.75\sigma_{\rm Fid}(S_8)$. When including smaller scales for $\ell<2500$, we find up to $-0.85\sigma_{\rm Fid}(S_8)$ shifts. However, at the baseline scale choice of $\ell_{\rm max} = 1500$, we observe a shift of only $-0.31\sigma_{\rm Fid}(S_8)$, well within our error budget.  
      
    \subsubsection{Multiplicative Bias}
      Multiplicative bias ($m$-bias) refers to a potential systematic in weak lensing that scales the shear field inferred from galaxy shapes measured in real images with respect to the true underlying lensing field. This is caused by the miscalibration of the galaxy shapes and shape noise bias \citep{2013MNRAS.429.2858M}. The simulation mocks are generated with one multiplicative bias value for each field; however, there are percent-level uncertainties on the multiplicative bias for HSC-Y1. We account for this by testing the pipeline with a ``true'' dataset a set of realisations generated with $m$-bias purposefully miscalibrated with $ \Delta m = \pm 0.01$, represented by $m$-bias+ and $m$-bias-  in Figure \ref{fig:scale cuts}. We find shifts of up to $0.33\sigma_{\rm Fid}(S_8)$ for our baseline scale cuts. It is worth noting that, in the case of the $\mathcal{B}$ mark function, we expect the impact of multiplicative bias can be predicted analytically and easily marginalised over (although we do not do so in this work).

\section{Results}\label{sec:results}
  We first present the constraints on $S_8$ found using our baseline choice inference (Section \ref{ssec: main result}). We then discuss the impact of the choice of mark on these results (Section \ref{ssec:comparing marks}). Finally, in Section \ref{ssec: cross v auto} we discuss the information in the cross- and auto-spectra of the marked field in relation to higher-order correlators.

  \subsection{Baseline Analysis}\label{ssec: main result}
    \begin{figure*}
      \centering
      \includegraphics[width=0.5\linewidth]{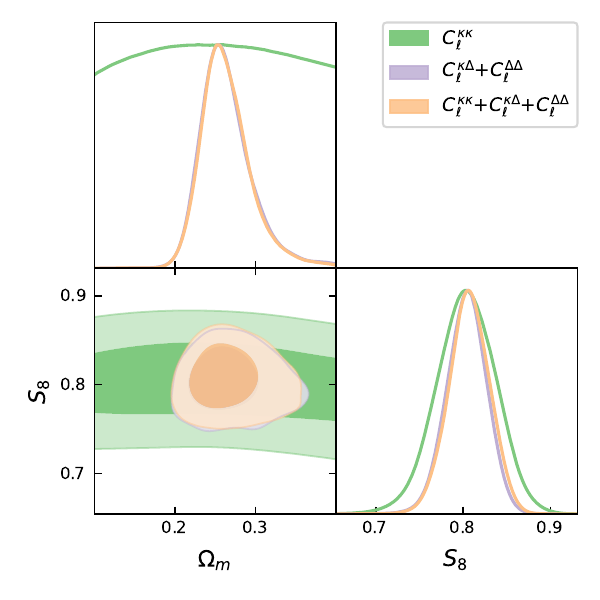}
      \caption{Constraints on $S_8$ and $\Omega_m$ using marked power spectra and HSC-Y1 data, using our baseline analysis choice of combining three mark functions ($\mathcal{A,B,C}$) over three smoothing scales of 2', 4', 10', using scale cuts of $\ell <1500$, and no cross correlations between separate tomographic bins, different mark functions, or smoothing scales. We colour the $68\%$ (inner line) and 95\% (outer line) confidence interval contours. $\kappa$ represents the original convergence field smoothed by 1', while $\Delta$ corresponds to the marked field, such that $C_\ell^{\kappa\kappa}$ corresponds to the typical power spectrum.}\label{fig:main_result}
    \end{figure*}

    \begin{figure*}
      \centering
      \includegraphics[width=\linewidth]{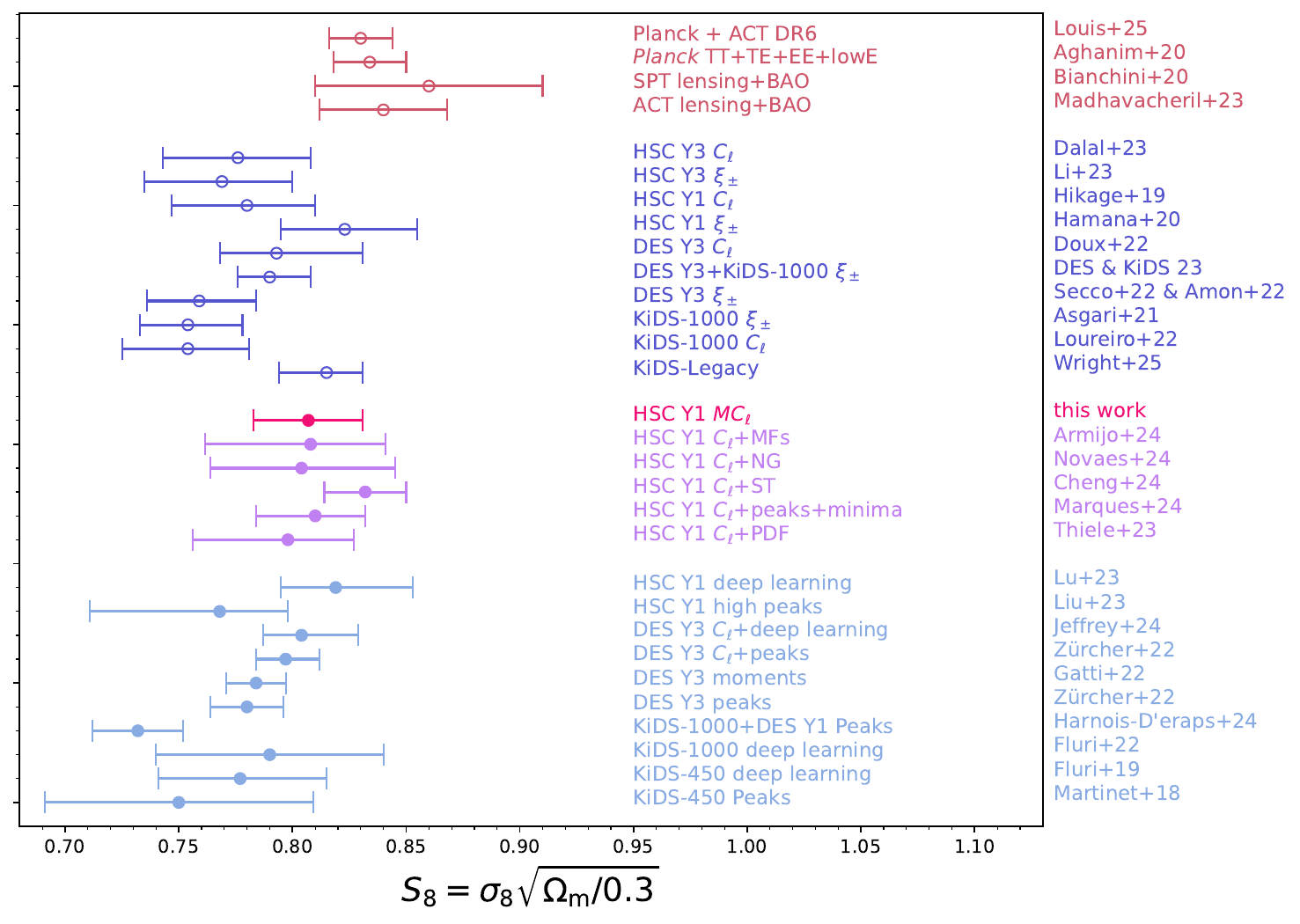}
      \caption{$S_8$ constraints from various surveys and methods. The top group in red corresponds to results from CMB surveys, meanwhile, the blue/purple are all from weak lensing. In purple we show the results of the other non-Gaussian (NG) statistics using HSC-Y1, with our result in pink. The bottom lighter blue shows constraints from other studies using NG statistics.}\label{fig:whisker plot}
    \end{figure*}
    For our baseline analysis, we combine all three mark functions and configurations with the scale cut $300 < \ell < 1500$ (Section \ref{ssec:scale_cuts}). Although the marked power spectra also contain information about the field on scales outside this range, which we discuss more in Section \ref{ssec: cross v auto}. We show results in Figure \ref{fig:main_result}, where the constraints found using the standard power spectrum, $C_\ell^{\kappa\kappa}$, are shown in green, the constraints obtained from the marked spectra ($C_\ell^{\kappa\Delta}+C_\ell^{\Delta\Delta}$) are shown in purple, and the combination of all spectra is shown in orange. The resulting constraints on $S_8$ are $S_8 = 0.803\pm 0.034$, $ S_8 = 0.804\pm 0.023$ and  $S_8 = 0.807\pm 0.024$ respectively.

    Our fiducial constraint using HSC-Y1 data is  $S_8 = 0.807\pm 0.024$, showing a 1.4$\times$ smaller error compared to that from $C_\ell^{\kappa\kappa}$ alone using the same scale cuts. We find adding   \( C_{\ell}^{\kappa \kappa}\) doesn't improve our constraint, likely due to the fact that the information in  $C_\ell^{\kappa\kappa}$ is already encoded in  \( C_{\ell}^{\Delta\Delta} \) (see discussions in Section \ref{ssec: cross v auto}). We also obtain $\Omega_{\rm m} = 0.265^{+0.022}_{-0.035}$. However, the inferred value of $\Omega_{\rm m}$ is affected by imperfect emulator accuracy (Section \ref{ssec:emulator}). Therefore, caution must be exercised before interpreting this result.

    Figure~\ref{fig:whisker plot} compares our result to other measurements of $S_8$. CMB-derived constraints appear at the top of the figure, followed by weak lensing results from two-point statistics, including both power spectra and correlation functions, shown in dark blue. Our result from marked spectra is shown in pink, while in purple we show results from other non-Gaussian analysis of HSC-Y1 in this series of papers. Finally, results from other studies using higher-order statistics are shown in light blue at the bottom. Overall, our constraint lies between previous weak lensing and CMB results, positioned near the centre of the so-called $S_8$ tension. The constraints are also in agreement with the most recent measurement from the KiDS Legacy data~\citep{2025kidslegacy}.

  \subsection{Comparison of Mark Functions}\label{ssec:comparing marks}
    \begin{figure*}
      \begin{subcaptionbox}{}[0.45\textwidth]
        {\includegraphics[width=\linewidth]{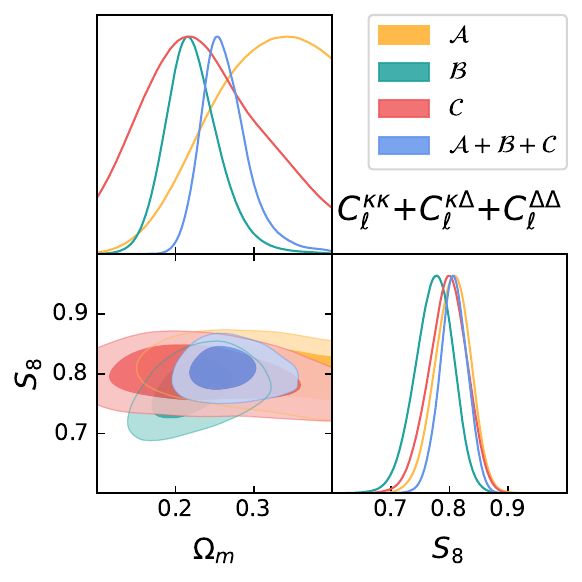}}
      \end{subcaptionbox}
      \centering
      \begin{subcaptionbox}{}[0.45\textwidth]
        {\includegraphics[width=\linewidth]{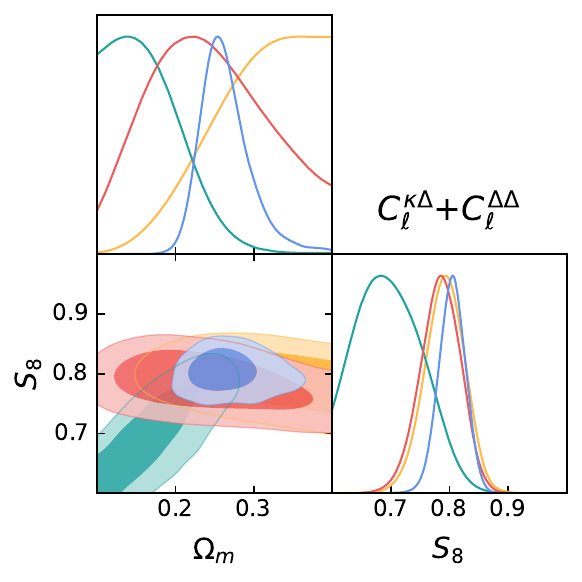}}
      \end{subcaptionbox}
      \begin{subcaptionbox}{}[0.45\textwidth]
        {\includegraphics[width=\linewidth]{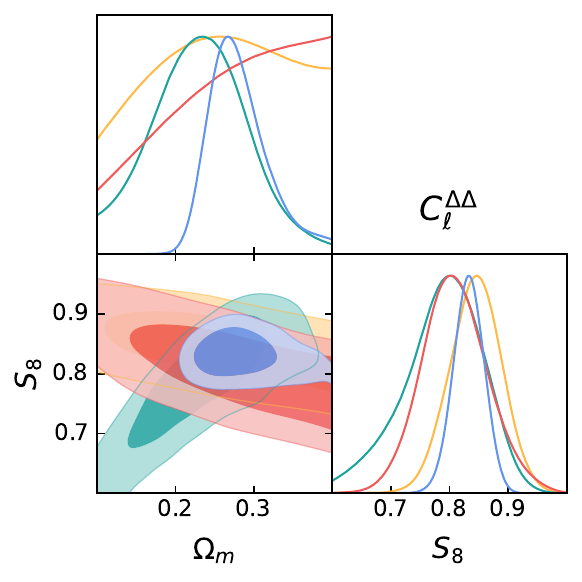}}
      \end{subcaptionbox}
      \begin{subcaptionbox}{}[0.45\textwidth]
        {\includegraphics[width=\linewidth]{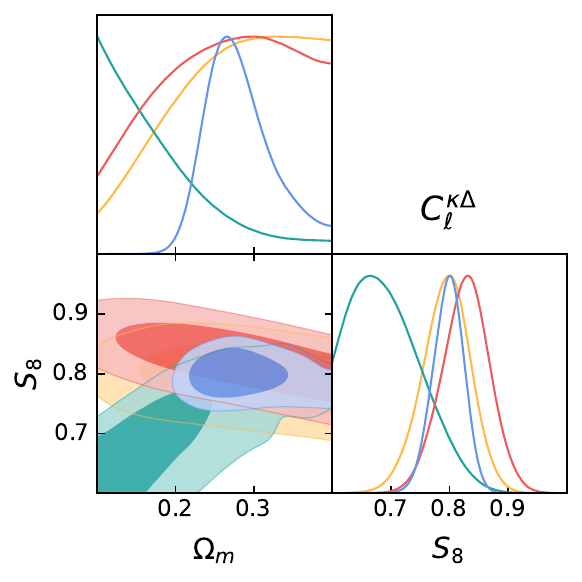}}
      \end{subcaptionbox}
      \caption{Constraints on $\Omega_{\rm m}$ and $S_8 = \sigma_8 \sqrt{\Omega_{\rm m}/0.3}$ from three different marked power spectra using HSC-Y1 data. $\Delta$ denotes the marked field, and $\kappa$ the original convergence field. The three marks emphasise different regions: $\mathcal{A}$ anti-correlates and up-weights medium over- and underdensities, $\mathcal{B}$ up-weights overdense regions, particularly the extremes of the field, and $\mathcal{C}$ up-weights only underdense regions. Contours show 68\% (inner) and 95\% (outer) confidence intervals. The blue contour represents the combined constraints from all three marks. Panel (a) includes all auto- and cross-spectra of both the original and marked fields. Panel (b) shows the combination of the auto and cross marked power spectra {\it without} the traditional power spectrum, while panels (c) and (d) isolate contributions from the auto and cross spectra, respectively. The uncertainty constraints from all displayed posteriors are displayed in Figure \ref{fig:barchart}.}\label{fig:comparing_marks}
    \end{figure*}

    \begin{figure*}
      \centering
      \includegraphics[width=0.9\linewidth]{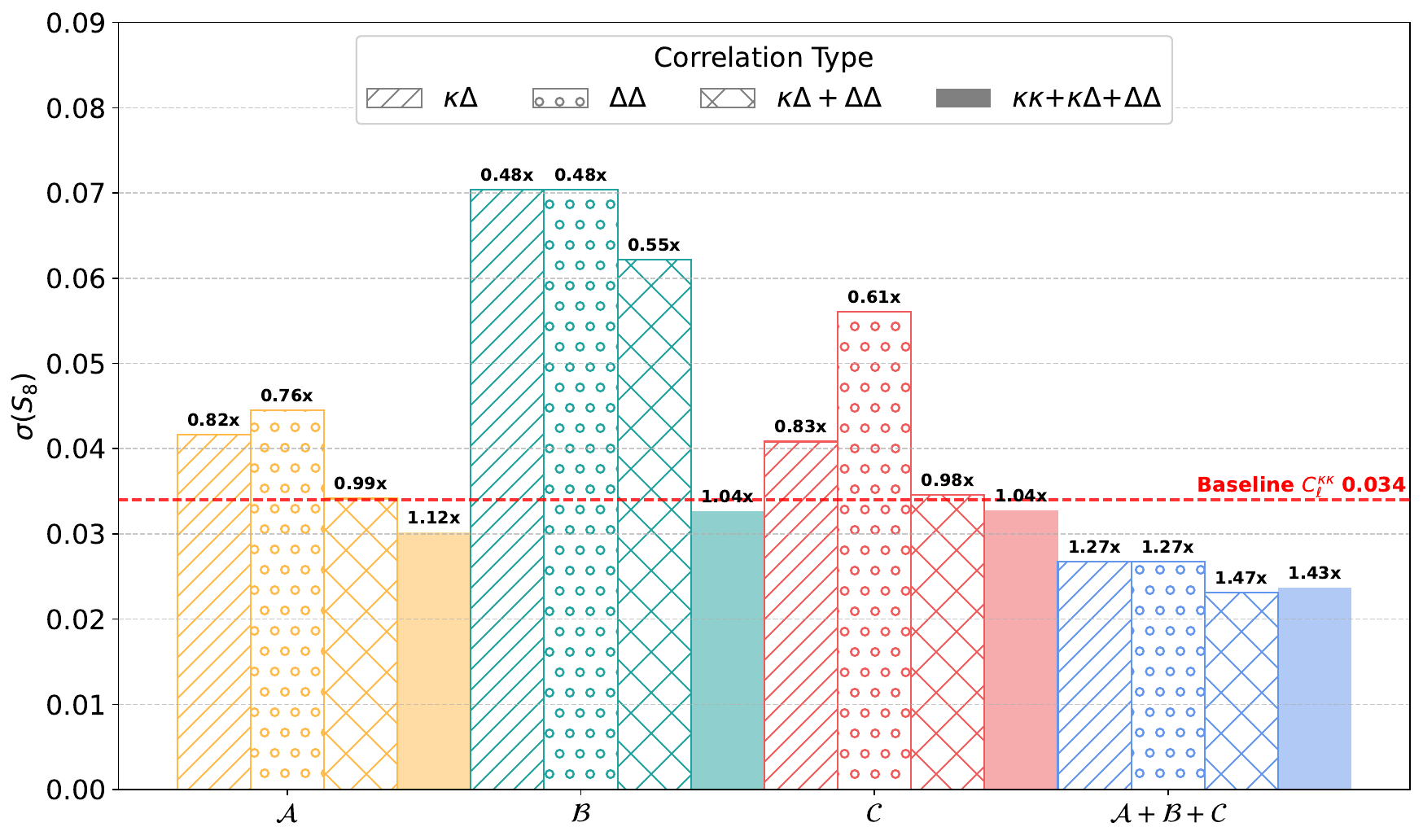}
      \caption{Posterior uncertainties on $S_8$ for different analysis configurations. Results are shown for individual mark choices ($\mathcal{A}$, $\mathcal{B}$, $\mathcal{C}$; in distinct colours) as well as their combination (blue). Within each configuration, we compare the constraints obtained from the correlation functions of the original convergence field $\kappa$ and the marked field $\Delta$. The baseline $S_8$ uncertainty from the auto-power spectrum of the unmarked field, $\kappa\kappa$, is indicated by the horizontal red line. Numbers above each bar denote the relative improvement with respect to this baseline.}\label{fig:barchart}
    \end{figure*}

    We next investigate the constraining power of the individual mark functions and its auto or cross spectra with the original field. As described in Section \ref{ssec:choice of mark}, the three marks have different properties, with $\mathcal{A}$ anti-weighting mild over- and underdensities, $\mathcal{B}$ multiplying by the smoothed field, up-weighting the more extreme values of the field, and $\mathcal{C}$ up-weighting underdensities.  

    We show the constraints obtained from HSC-Y1 data for each mark function, including all three smoothing scales, in Figure \ref{fig:comparing_marks}, showing the combination of all spectra in panel a, constraints without the power spectrum in b, and the auto and cross spectra in c and d, respectively. Mark $\mathcal{A}$ is shown in yellow,  $\mathcal{B}$ in red,  $\mathcal{C}$ in green, and the combination, representing our baseline analysis choice, in blue. We show the corresponding $S_8$ errors in Figure \ref{fig:barchart}, where the improvement in the value of $\sigma(S_8)$ found from $C^{\kappa\kappa}_{\ell}$ in each case is displayed in text above each bar. The horizontal red line shows the constraint from $C_{\ell}^{\kappa\kappa}$ at the same scale cuts, such that any result below it corresponds to an improvement.

    The results in this figure can be summarised as follows:
    \begin{itemize}
      \item From all three panels,  we see the combination of three marks significantly shrinks the contours, and no single mark function can obtain comparable constraints to the combination $\mathcal{A+B+C}$ shown in blue, especially with regard to $\Omega_{\rm m}$.
      \item Although in panel a), which corresponds to the solid bars in Figure \ref{fig:barchart}, the constraints on $S_8$ appear visually similar for each mark, $\mathcal{A}$, in yellow, has the best individual improvement factor of 1.12, compared to 1.04 for the others. Combining all three marks, however, leads to a factor of 1.43 improvement.
      \item For $\Omega_{\rm m}$ it is clear that $\mathcal{B}$  has the most constraining power,  with an error of $\sigma(\Omega_{\rm m})= 0.033$. Since $C_\ell^{\kappa\kappa}$ cannot constrain $\Omega_{\rm m}$ within our prior range, it is difficult to quantify the improvement factor in this case. In any case, the recovered error is around three times smaller than the width of the prior. We expect the leading information to come from some configurations of the bispectrum and trispectrum, which we discuss further in Section \ref{ssec: cross v auto}.
      \item In panels c) and d), corresponding to the hashed bars,  we can see the three marks have different degeneracy directions in the $S_8 - \Omega_{\rm m}$ plane. Most notably, $\mathcal{B}$ appears almost perpendicular to $\mathcal{C}$. This degeneracy-breaking could explain the constraining power of $\mathcal{B}$ in $\Omega_{\rm m}$, as well as why its combination with the other marks increases the overall constraining power.
      \item Even without including the standard power spectrum (i.e., $C_\ell^{\kappa\kappa}$), tight constraints on $S_8$ can still be achieved, as illustrated in Figure \ref{fig:barchart}. Notably, only the solid bar includes $\kappa\kappa$. However, significant improvement beyond the standard power spectrum is only seen when combining multiple mark functions, as shown by the blue curves. Additionally, the combination of the marked cross- and auto-spectra for $\mathcal{B}$ and $\mathcal{C}$ (dotted lines) yields constraints nearly as tight (an improvement factor of $\sim 0.99$ when compared with $C_\ell^{\kappa\kappa}$ alone.
    \end{itemize}
    Since each mark weights different regions of the convergence field, it is therefore somewhat intuitive that the constraints from combining these three mark functions are significantly enhanced, similar to how combining power spectra from different cosmological environments can yield tighter constraints \citep{Bonnaire_2022}. This complementarity, particularly in breaking parameter degeneracies, highlights the strength of the multi-mark approach and motivates its use in future analyses aimed at extracting maximal information from weak lensing data.

  \subsection{Relationship to n-point correlation functions}\label{ssec: cross v auto}
    The weak lensing bispectrum contains additional information compared to the power spectrum \citep{Takada_2004}, and success in tightening constraints of $S_8$ and $\Omega_{\rm m}$ has been found using third-order shear correlation functions or moments from other weak lensing surveys such as KiDS~\citep{Burger_2024} and DES \citep{Gomes_2025, Secco_2022}.
    \begin{figure*}
      \centering
      \begin{subcaptionbox}{\label{fig:left}}[0.45\textwidth]
        {\includegraphics[width=\linewidth]{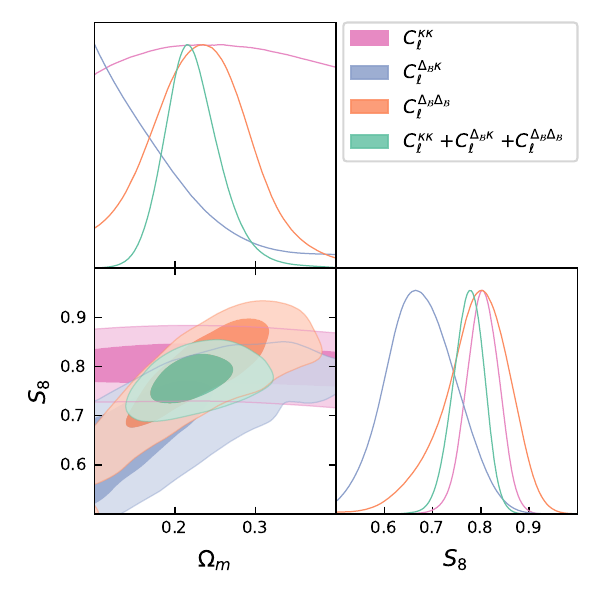}}
      \end{subcaptionbox}
      \begin{subcaptionbox}{\label{fig:right}}[0.45\textwidth]
        {\includegraphics[width=\linewidth]{figs/B_mark_cross_vs_auto.pdf}}
      \end{subcaptionbox}
      \caption{Posteriors of the different combinations of  $\mathcal{B}$ marked power spectra, using the mark function $m_\mathcal{B}=\kappa_\theta$ with the HSC-Y1 data. $\Delta$ denotes the marked field, while $\kappa$ is the original convergence field. We colour the 68\% (inner line) and 95\% (outer line) confidence intervals contours. On the left, we show the individual components, which are the power spectrum $C_\ell^{\kappa \kappa}$,  the cross spectra $C_\ell^{\Delta_\mathcal{B}\kappa}$ (bispectrum-like), and auto spectra $C_\ell^{\Delta_\mathcal{B}\Delta_\mathcal{B}}$ (trispectrum-like) of the marked fields.  On the right, we show the constraining power from different combinations of these terms.}\label{fig:cross_auto}
    \end{figure*}

    For the $m_\mathcal{A}$ and $m_{\mathcal{C}}$ marks the direct relation to n-point correlators is somewhat contrived, however in the case of the simplest mark, $m_\mathcal{B}=\kappa_\theta/\sigma(\kappa_\theta)$, the cross spectra $C_\ell^{\kappa \Delta_B}$ can be expressed purely in terms of the projected bispectrum, and the auto correlation $C_\ell^{\Delta_B \Delta_B}$ can be written as a four-point function.  In this sense, the mark choice of $m_\mathcal{B}$ is similar in spirit, to skew spectra \citep{Munshi_2020} or kurt spectra  \citep{Munshi_Lee_Dvorkin_McEwen_2022}, where one takes the correlation function of a polynomial of the field. The marked spectra differ in the introduction of the smoothing kernel, similar to the filtering step applied in \cite{harscouet_FSB}. 

    We show the parameter constraints derived from each of the spectra in Figure \ref{fig:cross_auto}. Panel a) focuses on the individual components, while panel b) shows the constraining power of different combinations. Focussing firstly on panel a), the ``bispectrum-like'' term, $C_\ell^{\kappa \Delta_B}$, is shown in the blue contour, while the auto-correlation $C_\ell^{\Delta_B \Delta_B}$, is shown in orange. The results obtained from the standard power spectrum $C_\ell^{\kappa \kappa}$ are shown in pink, and the combination of all three spectra is shown as a green contour. While each term has a slightly different $S_8-\Omega_{\rm m}$ degeneracy direction, we also see that the $C_\ell^{\kappa \Delta_B}$ contour hits the lower bound of the $\Omega_{\rm m}$ prior, and favours lower values of $S_8$, leading the auto-correlation to appear more constraining than the cross-correlation.

    In panel (b), we look at the combinations of the three spectra. We do not plot the $C_\ell^{\kappa \Delta_B}+C_\ell^{\Delta_B\Delta_B}$ combination, as it is not particularly constraining for this mark (see Figure \ref{fig:barchart}). It is interesting to note that the combination  $C_\ell^{\kappa\kappa} + C_\ell^{\Delta\Delta}$, plotted in purple, leads to better constraints than $C_\ell^{\kappa\kappa} + C_\ell^{\kappa\Delta}$, plotted in orange.

    One might expect that, since $C_\ell^{\kappa\Delta}$ is sourced by the bispectrum, it should dominate the improvement on cosmological constraints with respect to the power spectrum alone. This is because the bispectrum captures the leading-order non-Gaussian signal, has a higher signal-to-noise ratio than higher-order connected correlators (e.g. the trispectrum), and is less affected by observational noise.

    To understand this better, it is therefore useful to expand the marked power spectra in terms of n-point statistics. For simplicity we work in the flat sky approximation. Let $\Delta(\vec{x}) \equiv \kappa(\vec{x}) \kappa_\theta(\vec{x})$, where $\theta$ donates the smoothing scale\footnote{For simplicity we leave out the normalisation factor $\frac{1}{\sigma(\kappa_\theta)}$.}. From the convolution theorem, in Fourier space
    \begin{equation}
      \Delta_\vell =  \int \frac{d^2L}{(2\pi)^2} \kappa_{\theta,\vec{L}} \kappa_{\vell-\vec{L}}= \int \frac{d^2L}{(2\pi)^2}W(L\theta)\kappa_{\vec{L}}\kappa_{\vell-\vec{L}}
    \end{equation}
    where $W(L\theta)$ is the smoothing kernel. The cross spectrum of the original and marked fields is
    \begin{align}
      C_\ell^{\kappa\Delta}= \int \frac{d^2 {L}}{(2\pi)^2} \, W({L} \theta)\,B_\kappa({\ell}, {L}, |{\bf l} - {\bf L}|)
    \end{align}
    The window function restricts the effective contributions to modes with wavenumber  $L\lesssim1/\theta$, such that we get squeezed-like contributions of the form $B(\ell, 1/\theta, \ell-1/\theta)$.
    
    Similarly, the $\Delta$-$\Delta$ correlator can be expanded as 
    \begin{align}
      C_\ell^{\Delta\Delta} &= \Bigg[ 
      \underbrace{\int \frac{d^2L}{(2\pi)^2} W^2(L\theta)\,2C^{\kappa\kappa}_LC^{\kappa\kappa}_{|{\bf l}-{\bf L}|}}_{\text{Gaussian part}} \notag \\
      &\quad +  
      \underbrace{\int \frac{d^2L\,d^2L'}{(2\pi)^4} W(L\theta)W(L'\theta)\, T^{\kappa}({\bf L}, {\bf l}-{\bf L}, -{\bf L}',{-}{\bf l}+{\bf L}')}_{\text{non-Gaussian part}} \Bigg],
    \end{align}
    where $T^{\kappa}({\bf l}_1,{\bf l}_2,{\bf l}_3,{\bf l}_4)$ is the connected (non-Gaussian) trispectrum of the convergence field. We can see that, while the second term above is purely non-Gaussian, the first contribution contains only the Gaussian information found in the disconnected trispectrum. As in the case of $C_\ell^{\kappa\Delta}$, the range of scales the measurements are sensitive to depend on the smoothing scale $\theta$. Specifically, the Gaussian contribution above is sensitive to the power spectrum on scales $\ell\lesssim\ell_{\rm max}^{\rm 2pt}+1/\theta$, where $\ell_{\rm max}^{\rm 2pt}=1500$ is the scale cut used in our analysis. For a smoothing scale of e.g. $\theta=2'$, using $\ell\sim 1/\theta\sim1700$, this means that the Gaussian contribution to $C_\ell^{\Delta\Delta}$ is sensitive to the power spectrum on multipoles $\ell\lesssim 3200$. This explains the apparent additional constraining power from the marked power spectrum, likely coming from Gaussian fluctuations on smaller scales than available to the measured power spectrum, and not from the field's intrinsic non-Gaussianity. As a consistency test, we verified that the constraints obtained from $C_\ell^{\Delta\Delta}$ for larger smoothing scales were consistently less tight than those found for $\theta=2'$ (e.g. the error on $S_8$ grows by $121\%$ when using $\theta=10'$).

\section{Conclusions}\label{sec:conc}
 
  In this work, we present the first application of marked angular power spectra to weak lensing data, using simulated convergence maps tailored to the Subaru Hyper Suprime-Cam Year 1 (HSC-Y1) survey, which is the current survey with the most similar number density of galaxies to Stage-IV surveys such as the Rubin Observatory's Legacy Survey of Space and Time \citep[LSST,][]{2018arXiv180901669T} . We investigated the performance of three different mark functions, each designed to emphasise different density environments, by evaluating their statistical power in constraining cosmological parameters.

  To model the marked statistics as a function of cosmology, we used a Gaussian process emulator trained on a suite of 100 cosmological simulations varying $\Omega_{\rm m}$ and $\sigma_8$, each with 50 pseudo-independent realisations. The covariance matrix was estimated from 2268 pseudo-independent maps derived from 108 full-sky simulations at a fiducial cosmology, cut and masked to match the HSC-Y1 footprint and galaxy properties. We note that, for simplicity, we did not exploit cross-correlations between fields in different redshift bins, which could significantly tighten the constraints found using both power spectra and marked statistics.

  Our key findings are as follows:
  \begin{itemize}
    \item \textbf{The combination of different marked statistics enhances cosmological constraints (Figure~\ref{fig:comparing_marks}):} the combination of all three mark functions yields significantly tighter constraints on cosmological parameters compared to using individual marks or the standard power spectrum alone. This is particularly true for $S_8$ and $\Omega_{\rm m}$, with each mark probing different regions of the density field and contributing complementary information.
    \item \textbf{Improved $S_8$ constraints (Figures~\ref{fig:main_result} \&~\ref{fig:whisker plot}):} our fiducial analysis, which includes cross- and auto-correlations of marked fields with smoothing scales of 2, 4, and 10 arcminutes, and angular scales in the range $\ell \in [300,1500]$, yields a constraint of $S_8\equiv\sigma_8\sqrt{\Omega_{\rm m}/0.3} =  0.804\pm 0.023$, corresponding to an improvement factor of $\sim1.43$ (i.e. $32\%$ smaller uncertainties) over the constraint from $C_\ell^{\kappa\kappa}$ alone, under identical scale cuts.
    \item \textbf{Information leakage from small scales (Section \ref{ssec: cross v auto}):} due to the non-linear nature of the mark construction, marked power spectra exhibit a non-trivial mixing of scales, allowing small-scale information to leak into large-scale modes. We demonstrate that this effect can be mitigated through the choice of smoothing scale and validate our scale cuts ($\ell_{\max} = 1500$) against known systematics. Nevertheless, this implies that a significant fraction of the improvement to the constraining power can be brought about by Gaussian information on smaller scales, rather than purely non-Gaussian data combinations.
    \item \textbf{Robustness to systematics (Figure~\ref{fig:scale cuts}):} We assess the impact of key observational systematics, including baryonic effects, intrinsic alignments, multiplicative shear bias, and photometric redshift uncertainties, on the inferred cosmology. We find that, with appropriate smoothing choices and scale cuts, the bias in $S_8$ remains within $\sim0.4\sigma$ for all tested systematics.
    \item \textbf{Theoretical implications (Section~\ref{ssec: cross v auto}):} We show that the marked angular power spectra contain contributions analogous to bispectrum and trispectrum configurations, providing a practical and interpretable method of capturing non-Gaussian information without directly computing higher-order correlators.
  \end{itemize}
  These results demonstrate that marked power spectra, when applied to weak lensing data, are able to extract significant non-Gaussian information. Combined with the ease with which these statistics can be deployed in real data, this makes them a promising observable to exploit in ongoing and future experiments.

\section*{Acknowledgements}
  JAC is funded by a Kavli/IPMU PhD Studentship. JA is supported by JSPS KAKENHI Grant Number JP23K19064. DA acknowledges support from the Beecroft trust. MS is supported by JSPS KAKENHI Grant Numbers JP24H00215 and JP24H00221. LT is supported by JSPS KAKENHI Grant 24K22878. CPN thanks Instituto Serrapilheira for financial support. JL is supported by JSPS KAKENHI Grant Numbers 23H00107 and 25H00403. The Kavli IPMU is supported by the WPI (World Premier International Research Center) Initiative of the MEXT (Japanese Ministry of Education, Culture, Sports, Science and Technology). Numerical computations were [in part] carried out on the analysis servers and the general-purpose PC cluster at the Center for Computational Astrophysics, National Astronomical Observatory of Japan.

\section*{Data Availability}
  The code and data used for this analysis can be made available upon request to the authors.



\bibliographystyle{mnras}
\bibliography{bib} 



\appendix

\section{Emulator Bias}
\label{sec:emulator_appendix}
As briefly mentioned in Section \ref{ssec:emulator} where we discuss the tests performed on our emulator, when testing the full pipeline by inputting data vectors calculated from simulations with varied cosmology, our analysis shows a shift trend on the $\Omega_{\rm m}$ value recovered by the inference, visible in the right panel of Figure~\ref{fig:post}. It is worth noting that, in most of our analysis setups, $\Omega_{\rm m}$ is prior-dominated, and thus the metric used in this figure does not represent the emulator bias as a fraction of the real statistical uncertainties; a visual inspection of the contours indeed revealed that many of them are indeed prior-dominated, giving an artificially smaller standard deviation. Therefore, for posteriors that hit the prior bounds, we do not display any numerical offset. Based on this visual check,  we also mark cosmologies whose posteriors visually contain the true value within $1\sigma$ with a green circle. For those where the posterior appears truly biased, we show the numerical offset in terms of standard deviation beside the point. 

The offsets are significantly higher than those found for $S_8$, for which constraints are data-dominated. We find $m_\mathcal{B}$ to give the largest numerical contribution to this bias; however, it also drives the constraining power in $\Omega_{\rm m}$. The cosmology models with the worst bias also seem to be consistent across marks $m_\mathcal{A}$ and $m_\mathcal{B}$, but $m_\mathcal{C}$ has fewer trends toward overestimating $\Omega_{\rm m}$. We expect this mark to be no more sensitive to small-scale overdensities than the usual power spectrum, which suggests that the source of the discrepancy could be related to small scales; however, this trend is not seen clearly in the residuals of the power spectra themselves. We leave the investigation of the source of such bias to future work. For this reason, we place emphasis on the improved constraining power, rather than the value of $\Omega_{\rm m}$ inferred from our analysis.
\section{Simulation Discrepancies}\label{apsec:fiducial_vs_varied}
  \begin{figure*}
    \centering
    \includegraphics[width=\linewidth]{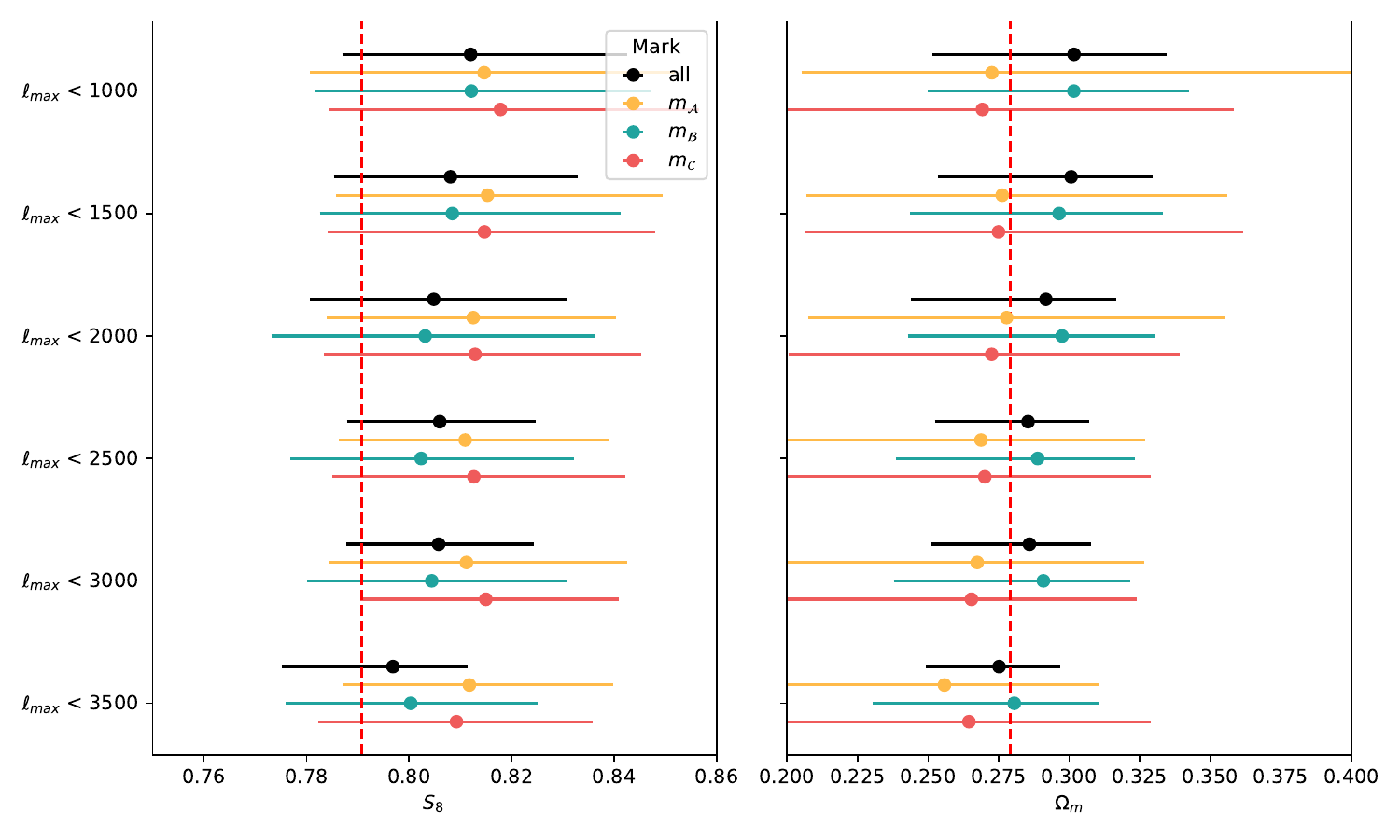}
    \caption{Posterior constraints for all three mark functions ($\mathcal{(A,B,C)}$ using smoothing scales of 2', 4', and 10' when inputting a data-vector calculated from the covariance-training simulation set, and evaluated using the emulator trained on the cosmo-varied simulations. The true cosmology of the covariance training set is marked in red. The three marks are shown in different colours, with the combination of the three shown in black. }
    \label{fig:fid_inf_scales}
  \end{figure*}
  We observe a systematic offset when inferring the cosmology of the covariance set simulations from \citep{Takahashi_2017}. Inference is performed using an emulator trained on the cosmo-varied suite, presented in \citep{marques2023cosmologyweaklensingpeaks}, based on the simulations of \citet{Shirasaki_2021}. This is a known phenomenon observed in the other NG HSC-Y1 papers using this analysis technique \citep{Cheng2025, Armijo:2024ujo,thiele2023cosmologicalconstraintshscy1, grandon2024impactbaryonicfeedbackhsc, novaes2024cosmologyhscy1weak, marques2023cosmologyweaklensingpeaks}.

Figure~\ref{fig:fid_inf_scales} shows the combined posterior constraints for all three mark functions across the three baseline smoothing scales, when inputting the mean of the covariance training simulations as a data vector. In black, we plot the combination of all marks, as used in the baseline analysis. In colour, we plot the results using only one of the mark fields such that the data vector is $[C_\ell^{\kappa\kappa},C_\ell^{\kappa\Delta},C_\ell^{\Delta \Delta}]$. For all analysis choices, $S_8$ is overestimated, regardless of the choice of $\ell_{\max}$, while $\Omega_{\rm m}$ trend varies depending on mark choice. The combined inference remains largely consistent with expectations, though a systematic shift in $S_8$ persists. 
\begin{figure*}
    \centering
    \includegraphics[width=\textwidth]{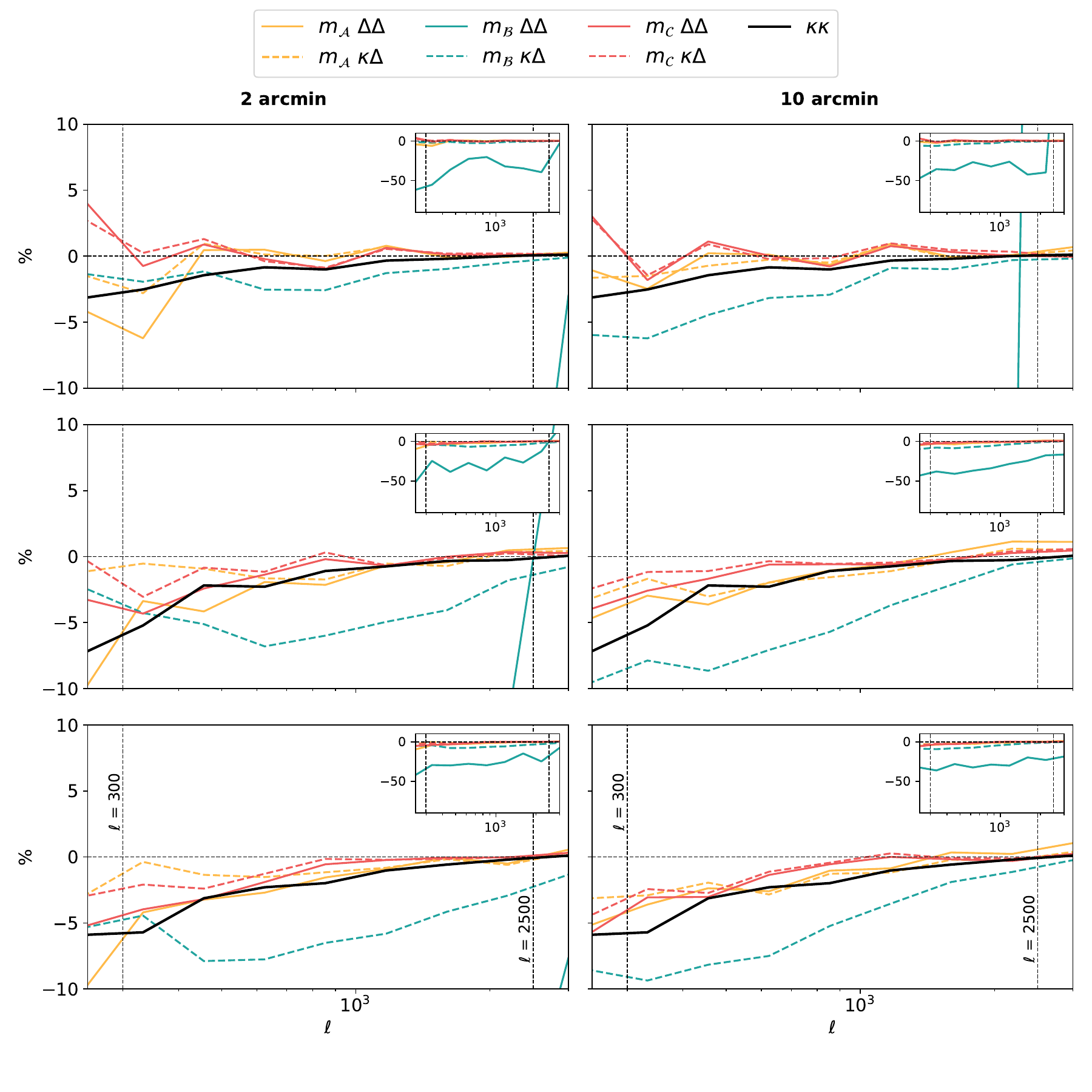}
    \caption{Percentage residuals between the power spectra constructed by the emulator trained onthe varied cosmology simulations, and true power spectra measured from the covariance training simulations directly. The residuals are defined as $100 \times (C_\ell^{\rm true} - C_\ell^{\rm emu}) / C_\ell^{\rm true}$, for the covariance training set. We show the residuals for all three mark functions, both cross- and auto-spectra, and two smoothing scales: 2' and 10'. Results are shown across all redshift bins. The standard $C_\ell^{\kappa\kappa}$ spectrum is shown in black.}
    \label{fig:fid_vs_varied}
\end{figure*}
This bias is likely caused by a mismatch in the predicted and true power spectra. It is known that for the unmarked power spectrum $C_\ell^{\kappa \kappa}$, the cosmo-varied suite has less power than the covariance suite on large physical scales. This effect has been confirmed in prior analyses, and is discussed in the Appendix of \citep{Cheng2025}. We explore the difference at the power spectrum level in Figure~\ref{fig:fid_vs_varied}, which shows the residuals between the emulated and true marked power spectra, broken down by mark type, smoothing scale, and redshift bin.  

In black, labelled $\kappa \kappa$, we show the standard power spectrum $C_\ell^{\kappa \kappa}$ , where the lack of power on large scales is visible. The three mark functions exhibit distinct behaviours:
\begin{itemize}
    \item The $m_\mathcal{A}$ mark (yellow), which up-weights and anti-correlates under- and overdense regions, shows moderate, scale-dependent deviations, which appear to worsen with increasing redshift.
    \item The $m_\mathcal{B}$ mark (blue), which overweights overdensities, displays the largest residuals, particularly in the cross-spectrum component shown in the inset, with deviations an order of magnitude larger than other marks. This term is sensitive to some configurations of the squeezed bispectrum.  This spectrum in particular, has very low signal-to-noise, with a large variance across the covariance simulations. It is possible that the 50 realisations of the cosmo-varied simulations are not enough to accurately predict this spectrum. Alternatively, unlike the auto spectrum, $C^{\kappa\Delta}_\ell$ does not contain information proportional to the two-point function, and at all scales captures small scale information on the scale of the smoothing scale. However, we see a similar behaviour for 10 arcmin and 2 arcminutes, suggesting this behaviour is more complicated than just a small-scale effect. A future investigation using the bispectrum applied to this dataset could help illuminate this discrepancy.
    \item The $m_\mathcal{C}$ mark (red), which up-weights underdense regions, shows smoother residuals and is less strongly biased. Interestingly, at lower redshifts it appears to have positive residuals, which switch with increasing redshift.
  \end{itemize}

  Interestingly, while $m_\mathcal{B}$ shows the largest spectral residuals, this does not translate into a proportionally large bias in parameter inference (see Figure~\ref{fig:fid_inf_scales}). This suggests that some of the deviation may cancel out in the compressed data vector, or that the source of the discrepancy could be the low signal to noise, which is accounted for by the covariance matrix.

  The cause of the discrepancy between the two simulations is not fully understood. The covariance simulations have been validated against HALOFIT \citep{Smith_2003} predictions and reproduce the convergence power spectrum within $5\%$ up to $\ell \sim 3000$, however they are almost affected by finite thickness effects, making their lensing power spectra very large scales smaller than the popular halofit predictions \citep{Takahashi_2017}. The finite thickness effects are totally simulation-dependent and cannot be corrected with any post processing.  The cosmo-varied simulations from \citet{Shirasaki_2021} have higher mass resolution and are expected to contain more small-scale power. However, the observed deficit in large-scale power is harder to explain. One possibility is the use of adaptive box sizes along the light cone in the cosmo-varied suite, which may introduce subtle inconsistencies across scales.

  These findings highlight the importance of consistency between simulation suites when building emulators for non-Gaussian statistics, especially when scale mixing is involved. Further investigation into the discrepancy between the simulation sets is needed to fully interpret these biases, and it may be difficult to find perfect consistency of $S_8$ inferences among different mocks at this stage.

\section{Impact of Smoothing and Scale Cuts on Systematic Bias}
\label{apec:systematics}
\begin{figure*}
    \centering
    \includegraphics[width=\linewidth]{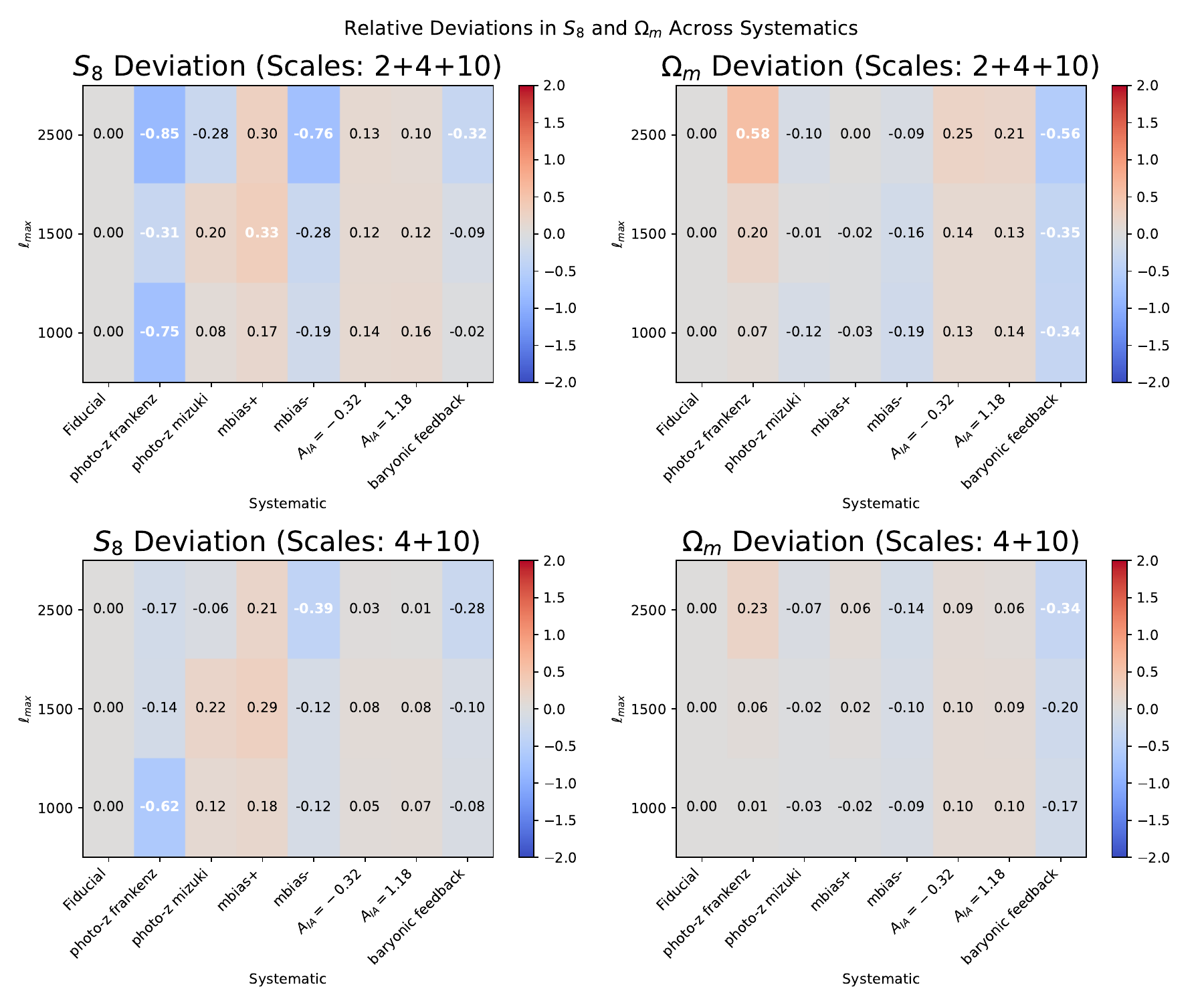}
    \caption{Heatmap displaying the bias introduced by performing inference on contaminated data vecotrs, compared to the true cosmology. The color indicates the deviation in the inferred parameters $S_8$ and $\Omega_{\rm m}$, expressed in units of their standard deviation $\Delta \sigma$. Red denotes overestimation, blue underestimation, and white text highlights shifts larger than 0.3$\sigma$. The top row corresponds to the baseline analysis choices of smoothing scales, while the bottom row removes the 2 arcminute mark functions.  }
    \label{fig:sys_heatmap}
\end{figure*}
{One might expect that including smaller physical scales would amplify the bias of systematic effects not modelled by the simulations. However, as discussed in Section~\ref{ssec: cross v auto}, applying scale cuts to the marked power spectra does not fully isolate the information from these scales, due to the intrinsic scale mixing introduced by the mark function. More specifically, the smoothing scale used in the mark determines the degree of ``leakage'' from smaller physical scales, such that information from the smoothing scale size could always be present in the data.
In Figure~\ref{fig:sys_heatmap}, we illustrate how the inferred cosmological parameters shift in response to different systematic effects as a function of $\ell_{\max}$. We examine three smoothing scales used in the baseline analysis, $2',  4'$ and $ 10'$, and also compare to the case where the $2'$ smoothing is excluded. Each column corresponds to a different systematic, while the $y$-axis denotes increasing $\ell_{\max}$ (with $\ell_{\min} = 300$ fixed). 
The colour indicates the deviation in the inferred $S_8$ and $\Omega_{\rm m}$ from their fiducial values, in units of their standard deviation $\Delta\sigma$. Red denotes an overestimation, blue an underestimation, and white text flags deviations greater than $0.3\sigma$.
For baryonic effects, we observe the expected behaviour increasing suppression of $S_8$ as smaller scales are included.

More surprising trends appear for other systematics. For example, the model with strong intrinsic alignment ($A_{IA} = 1.18$, shows increasing bias with \emph{larger} scale cuts -- contrary to the standard expectation that smaller scales drive stronger systematics. Similarly, significant bias is observed when using the \textsc{frankenz} photo-$z$ estimator, even with the $2'$ scale removed. An important caveat to this figure is that this illustrates the behaviour from all three marks, corresponding to our baseline analysis. The behaviour of systematics for each mark also varies, such that statements true of systematic bias for this analysis may not be true in the general case of `marked statistics'.

These trends highlight the complex interactions between mark choice, smoothing scale, and systematics. Since the marked statistics mix scales and amplify non-Gaussian information, they may respond to systematics in less predictable ways than standard two-point functions. A detailed analysis of these interactions is left to future work.}

\section{Scale Dependence on HSC-Y1 Data}\label{apsec:S_8scale}
  \begin{figure*}
    \centering
    \includegraphics[width=\linewidth]{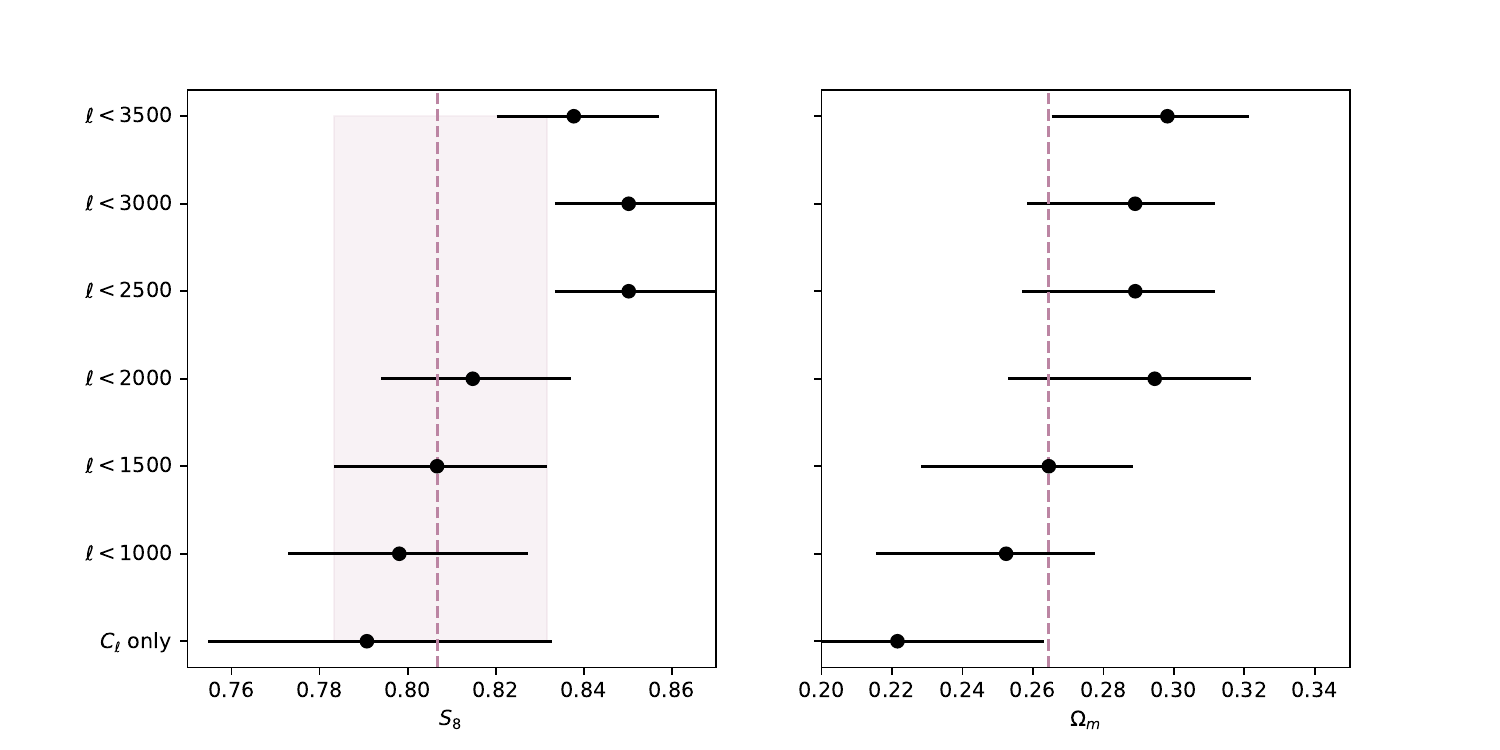}
    \caption{\
    Constraints on $S_8$ and $\Omega_m$ from HSC-Y1 data using the baseline analysis choices of three marks and three smoothing scales, but varying the scale cuts with $\ell_{min}=300$ fixed.. We do not take cross-correlations between different mark models or tomographic bins. We show the constraint from our baseline analysis $(\ell<1500)$, which passed systematic validation for $S_8$ in pink. }
    \label{fig:sims_vs_data_S_8trend}
  \end{figure*}
  In the analysis of the real HSC-Y1 data, we observe a consistent increase in the inferred value of $S_8$ as smaller angular scales are included in the data vector. We show the constraints on HSC-Y1 data in Figure~\ref{fig:sims_vs_data_S_8trend}, where $S_8$ rises as higher $\ell$ modes are added. This trend is not seen when repeating the same analysis on simulated data vectors, as shown in Figure \ref{fig:fid_inf_scales}, but is true for both the marked and standard power spectra. In simulations, $S_8$ estimates remain either consistent or exhibit a weak decrease with increasing $\ell_{\max}$. This is true whether the data vector is generated from the covariance suite or from the cosmology-varied simulations.  

  This discrepancy raises several possible explanations. One is that residual systematics in the HSC-Y1 data, such as unaccounted baryonic effects, intrinsic alignment, or photo-$z$ shifts, could contribute to this scale-dependent bias. However, extensive systematic testing (see Section~\ref{apec:systematics}) suggests that such effects alone cannot fully account for the observed trend. Another possibility is that the real Universe contains small-scale structures or physics not captured by the current simulations, which rely on gravity-only (or fixed feedback) models. More likely is that the simulations do not have high enough resolution at small scales to accurately model the HSC-Y1 data, leading to a lack of power on small scales. This would cause the real universe to appear to have a higher clustering at smaller scales, boosting the inferred value of $S_8$.{ It is also known that the high convergence regions, which are highly up-weighted by mark $\mathcal{B}$, contain rich information about cluster-sized halos,  however the simulations we use likely cannot accurately predict this regimem leading to a significant bias.  } Nevertheless, our fiducial scale cuts ($\ell_{\rm max}=1500$) ensure that this potential bias is kept within the statistical uncertainties. We leave a detailed investigation of this behaviour to future work.


\bsp	
\label{lastpage}
\end{document}